\documentclass{aastex63}

\received{December 16 2019}
\accepted{February 16 2021}
\submitjournal{AJ}

\begin{document}

\title{Deep Contrast and Companion Detection Using the EvWaCo Testbed Equipped with an Achromatic Focal Plane Mask and an Adjustable Inner Working Angle}

\author{Mary Angelie Alagao}
\affil{National Astronomical Research Institute of 
Thailand \\
260 Moo 4, T. Donkaew, A.  Maerim \\
Chiang Mai, Thailand, 50180}
\affil{PhD Program in Astronomy,  
Department of Physics and Materials Science \\
Faculty of Science, Chiang Mai University \\
Chiang Mai, Thailand, 50200}

\author{Adithep Kawinkij}
\affiliation{National Astronomical Research Institute of Thailand \\
	260 Moo 4, T. Donkaew, A.  Maerim \\
	Chiang Mai, Thailand, 50180}

\author{Christophe Buisset}
\affiliation{National Astronomical Research Institute of Thailand \\
260 Moo 4, T. Donkaew, A.  Maerim \\
Chiang Mai, Thailand, 50180}

\author{Apirat Prasit}
\affiliation{National Astronomical Research Institute of Thailand \\
	260 Moo 4, T. Donkaew, A.  Maerim \\
	Chiang Mai, Thailand, 50180}

\author{Thierry Lepine}
\affiliation{Institut d'Optique Graduate School \\
    18 Rue Benoit Lauras \\
	Saint-Etienne, France 42000}
\affiliation{Hubert Curien Laboratory (CNRS, UMR 5516)
	18 Rue Benoit Lauras \\
	Saint-Etienne, France 42000}

\author{Yves Rabbia}
\affiliation{Universit\'{e} de la C\^{o}te d' Azur, OCA, CNRS\\
Lagrange, Boulevard de l’Observatoire \\
B.P. 4229 F-06304 NICE Cedex-43, France}

\author{Anthony Berdeu}
\affiliation{National Astronomical Research Institute of Thailand \\
	260 Moo 4, T. Donkaew, A.  Maerim \\
	Chiang Mai, Thailand, 50180}
\affiliation{Department of Physics, Faculty of Science, Chulalongkorn University\\
254 Phayathai Road, Pathumwan\\
Bangkok, Thailand 10330}

\author{Eric Thiebaut}
\affiliation{Univ Lyon, Univ Lyon1, Ens de Lyon, CNRS, Centre de Recherche Astrophysique de Lyon UMR5574, F-69230, Saint-Genis-Laval, France}

\author{Maud Langlois}
\affiliation{Univ Lyon, Univ Lyon1, Ens de Lyon, CNRS, Centre de Recherche Astrophysique de Lyon UMR5574, F-69230, Saint-Genis-Laval, France}

\author{Michel Tallon}
\affiliation{Univ Lyon, Univ Lyon1, Ens de Lyon, CNRS, Centre de Recherche Astrophysique de Lyon UMR5574, F-69230, Saint-Genis-Laval, France}

\author{Supachai Awiphan}
\affil{National Astronomical Research Institute of Thailand \\
	260 Moo 4, T. Donkaew, A.  Maerim \\
	Chiang Mai, Thailand, 50180}

\author{Eugene Semenko}
\affil{National Astronomical Research Institute of Thailand \\
260 Moo 4, T. Donkaew, A.  Maerim \\
Chiang Mai, Thailand, 50180}

\author{Pakakaew Rittipruk}
\affil{National Astronomical Research Institute of Thailand \\
	260 Moo 4, T. Donkaew, A.  Maerim \\
	Chiang Mai, Thailand, 50180}

\author{David Mkrtichian}
\affil{National Astronomical Research Institute of Thailand \\
	260 Moo 4, T. Donkaew, A.  Maerim \\
	Chiang Mai, Thailand, 50180}

\author{Apichat Leckngam}
\affil{National Astronomical Research Institute of Thailand \\
	260 Moo 4, T. Donkaew, A.  Maerim \\
	Chiang Mai, Thailand, 50180}

\author{Griangsak Thuammasorn}
\affil{National Astronomical Research Institute of Thailand \\
	260 Moo 4, T. Donkaew, A.  Maerim \\
	Chiang Mai, Thailand, 50180}

\author{Pimol Kaewsamoeta}
\affil{National Astronomical Research Institute of Thailand \\
	260 Moo 4, T. Donkaew, A.  Maerim \\
	Chiang Mai, Thailand, 50180}

\author{Anuphong Inpan}
\affil{National Astronomical Research Institute of Thailand \\
	260 Moo 4, T. Donkaew, A.  Maerim \\
	Chiang Mai, Thailand, 50180}

\author{Teerawat Kuha}
\affil{National Astronomical Research Institute of Thailand \\
	260 Moo 4, T. Donkaew, A.  Maerim \\
	Chiang Mai, Thailand, 50180}

\author{Auychai Laoyang}
\affil{National Astronomical Research Institute of Thailand \\
	260 Moo 4, T. Donkaew, A.  Maerim \\
	Chiang Mai, Thailand, 50180}

\author{Worawat Somboonchai}
\affil{National Astronomical Research Institute of Thailand \\
	260 Moo 4, T. Donkaew, A.  Maerim \\
	Chiang Mai, Thailand, 50180}

\author{Suchinno Kanthum}
\affil{National Astronomical Research Institute of Thailand \\
	260 Moo 4, T. Donkaew, A.  Maerim \\
	Chiang Mai, Thailand, 50180}

\author{Saran Poshyachinda}
\affil{National Astronomical Research Institute of 
	Thailand \\
	260 Moo 4, T. Donkaew, A.  Maerim \\
	Chiang Mai, Thailand, 50180}

\author{Boonrucksar Soonthornthum}
\affil{National Astronomical Research Institute of 
Thailand \\
260 Moo 4, T. Donkaew, A.  Maerim \\
Chiang Mai, Thailand, 50180}



\begin{abstract}
The evanescent wave coronagraph uses the principle of frustrated total internal reflection (FTIR) to suppress the light coming from the star and study its close environment. Its focal plane mask is composed of a lens and a prism placed in contact with each other to produce the coronagraphic effect. In this paper, we present the experimental results obtained using an upgraded focal plane mask of the Evanescent Wave Coronagraph (EvWaCo). These experimental results are also compared to the theoretical performance of the coronagraph obtained through simulations. Experimentally, we reach a raw contrast equal to a few  $10^{-4}$ at a distance equal to 3  $\lambda/D$ over the full I-band $(\lambda_c =$ 800 nm, $\Delta\lambda/\lambda \approx 20\%)$ and equal to 4  $\lambda/D$ over the full R-band $(\lambda_c =$ 650 nm, $\Delta\lambda/\lambda \approx 23\%)$ in unpolarized light.
However, our simulations show a raw contrast close to $10^{-4}$ over the full I-band and R-band at the same distance, thus, confirming the theoretical achromatic advantage of the coronagraph.  We also verify the stability of the mask through a series of contrast measurements over a period of 8 months. Furthermore, we measure the sensitivity of the coronagraph to the lateral and longitudinal misalignment of the focal plane mask, and to the lateral misalignment of the Lyot stop.
\end{abstract}




\section*{}\vspace{-1.5cm}

\section{Introduction} \label{sec:intro}

High-contrast imaging of exoplanets is essential in gaining a better understanding of the formation and evolution of planetary systems. With enough signal from the planet, high-contrast imaging reveals not only the spectro-photometry of its atmosphere but also the different interactions in the close environment of the star \citep{Milli}\citep{Traub}\citep{Mawet}. However, direct imaging of exoplanets is a highly challenging goal because of the very high contrast and the very small angular separation existing between the planet and its parent star. In fact, contrasts are ranging from $C \approx 10^{-3}$ for hot giant planets in the infrared to $C \approx 10^{-10}$ for Earth-like planets in the visible \citep{Mawet}, and a contrast between $10^{-6}$ and $10^{-7}$ in the near-IR is required to image warm Jupiter-like planets \citep{Burrows}. The typical angular separation is equal to 0.1 arc second for Earth-like planets and 1 arc second for Jupiter-like planets, pertaining to a distance of 1 AU, seen at 10 pc.

Regarding those two constraints, instrumental issues such as diffraction and scattering have serious adverse effects on the detection. To be able to measure such contrasts, it is required to eliminate diffraction effects and to limit (and control) wavefront errors. 

The most recent ground-based coronagraphs installed on 8-10 meter class telescopes combined a high contrast performance coronagraph and a high-order adaptive optics to simultaneously i) mask the central object, ii) eliminate the diffraction effects, and iii) control the wavefront errors. For example, the Spectro-Polarimetric High-Contrast Exoplanet Research (SPHERE) installed on the VLT obtained a contrast of a few $10^{-7}$ for a distance between 6 $\lambda/D$ (0.3'') and 30 $\lambda/D$ (1.6'')  from the optical axis for a wavelength of 2.11 µm (K-band). These numerical figures are based on the results of the high contrast imaging of Sirius A \citep{Vigan}\citep{Beuzit}. Similarly, the measured 5$\sigma$ contrasts during the first light of the Gemini Planet Imager (GPI), which is installed on the Gemini South Telescope in Chile, were $10^{-6}$ at 17 $\lambda/D$ (1.2'') and $10^{-5}$ at 8 $\lambda/D$ (0.3'')  in the H-band $(\lambda_{mid}$ = 1.6 µm) \citep{Macintosh}. In this framework, the aim of the EvWaCo project is to develop a new kind of simple and cost-effective coronagraph for ground-based telescopes with state-of-the-art performance comparable to the ones in SPHERE and GPI.  

The National Astronomical Research Institute of Thailand (NARIT) has developed a testbed to demonstrate in laboratory conditions the capability of the Evanescent Wave Coronagraph (EvWaCo) to reach comparable results at optical wavelengths \citep{Buisset}\citep{Alagao}. The suppression of the star light in EvWaCo is achieved by placing in contact a prism and a lens at the telescope focal plane. The prism is oriented at an angle of incidence such that light should undergo total internal reflection. However, when a glass medium, such as a lens, is placed in contact at the focus, the beam is transmitted by means of captured evanescent waves and this phenomenon is known as “Frustrated Total Internal Reflection (FTIR)” \citep{Rabbia}\citep{Zhu}\citep{Gross}. In the case of an off-axis beam, the beam undergoes total internal reflection as soon as an air gap is met. This implies that with proper shaping of the materials in contact, the rejection profile can be adjusted. Furthermore, from the properties of the evanescent waves, we can theoretically show that the size of the mask inherently adapts itself to the wavelength, that is, the longer the wavelength, the larger the rejection profile \citep{Buisset}\citep{Alagao}. With these capabilities, we expect that the EvWaCo focal plane mask will provide an achromatic rejection. 

In its current form, the EvWaCo testbed comprises off-the-shelf components mounted on a passive support. This setup operates at ambient temperature without an adaptive optics setup to correct for the wavefront distortions induced by the optical quality of the components and by inhomogeneities of the air refractive index. In our previous paper \citep{Buisset1}, we presented the performance of EvWaCo in unpolarized broadband light over the full I-band (700 nm - 900 nm) with a central wavelength $\lambda_c$ = 800 nm and of spectral ratio $\Delta\lambda/\lambda_c \approx 25\%$. We used a large Lyot stop of diameter equal to 78\% of the aperture stop, thus, yielding a throughput close to 60\%. 

The measured contrast performance was a few $10^{-6}$ at distances from the optical axis in the focal plane between 10 $\lambda/D$ and 20 $\lambda/D$, and a few $10^{-7}$ between 20 $\lambda/D$ and 50 $\lambda/D$. The Inner Working Angle (IWA) was close to 6 Airy radii along the x-axis, and 8 Airy radii along the y-axis. We have also demonstrated that the signal obtained from a companion 30,000 times fainter than the star, located at a distance equal to 15 $\lambda/D$ from the center of the star image can be detected with an SNR equal to 75 \citep{Buisset1}. 

\section{The Evanescent Wave Coronagraph Project} \label{sec:EvWaCo}

These results lay the groundwork for the formation of the EvWaCo Project. The EvWaCo project is part of the Thai Space Consortium that was established in 2018 and aims at developing the human capacities, facilities and technology for space observations and exploration in Thailand. In this framework, the EvWaCo project aims at developing state-of-the-art technologies for high contrast observation with the medium and long term objective to provide sub-systems (coronagraphs, focal plane masks, masks, adaptive optics or spectrographs) for future space missions dedicated to direct detection and characterization of exoplanets.   

Under this project, an EvWaCo prototype equipped with an adaptive optics system will be installed on the 2.4 m Thai National Telescope at horizon 2022. The prototype is currently under development at the NARIT Center for Optics and Photonics with the support of Centre de Recherche Astrophysique de Lyon (CRAL), Institut d'Optique Graduate School (IOGS) and Laboratoire Hubert Curien (LabHC). It is specified to observe in the R- and I-bands with a raw contrast of $10^{-4}$ at an inner working angle of 0.5". The limiting magnitude of the central object is 8. This coronagraph is set to observe the science cases briefly discussed hereafter.

The coronagraph will be used to directly image the circumstellar disks, and other fine structures hidden in the bright stellar light. Among the suitable targets are the young stellar objects and pre-main sequence stars \citep{Watson}\citep{Tamura}. Another would be the direct imaging of brown dwarf binaries and M-dwarfs. The brown dwarf binary HD 130948 BC was detected in the I-band using the lucky imaging instrument, FastCam, installed at the 2.5 m Nordic telescope with an $SNR \approx 9$. The binary stars were separated by 2.561'' $\pm$ 0.007'' (46.5 AU) with a magnitude of $\Delta I \approx$ 11.30 ± 0.11(C $\approx$ $3.10^{-5}$) \citep{Labadie}. In the same spectral band, high resolution images  of 490 mid- to late- M dwarfs were also obtained using the FastCam installed at a 1.5 m telescope \citep{Cortes-Contreras}. These detections have been obtained without a coronagraph. Using a coronagraph would significantly increase the capability of detecting brown dwarf binaries and M-dwarfs closer to the star. In addition, these targets were detected at the current observing wavelength of the EvWaCo.

Moreover, taking advantage of EvWaCo's capability to collect the light coming from both the star and the companion, the coronagraph will be used to validate the candidate host stars with stellar companion or chance-aligned background star in Kepler \citep{Coughlin, Thompson}, K2 \citep{Yu, Mayo}, and TESS \citep{Stassun1, Stassun2} surveys. Many of the transit-like signals are false positives \citep{Morton}. The false positive scenarios are considered to be blended and unblended eclipsing binary stars, where the eclipsing binary may be physically associated with primary target or maybe a chance-alignment with the primary target. Follow-up photometric observations can be done to validate the contaminating sources.

The aim of this paper is to present both the experimental results and the simulations of the coronagraph. The experimental results are obtained using the new achromatic focal plane mask (FPM) with an adjustable inner working angle. The focal plane mask comprises a dedicated opto-mechanical support designed to adjust the contact between the lens and the prism, thus modifying the shape of the mask and its rejection profile.

This paper is outlined as follows. We present an overview of the EvWaCo testbed in Section 3. This testbed has been upgraded to accommodate the capabilities of the new focal plane mask. Section 4 presents mask reflection at varying pressure, and the experimental and simulation results for the mask reflection at two wavelengths: $\lambda_1$ = 800 nm and $\lambda_2$ = 650 nm. We also present the pupil irradiance distribution in this section. The raw contrast calculation is presented in Section 5. We also illustrate the contrast performance of EvWaCo by showing the gain provided by this coronagraph for the observation of a companion 25000 times fainter than the star and located at a distance equal to 5.5 $\lambda/D$ with an SNR equal to 11. In Section 6, we present the repeatability of the coronagraph performance in terms of the contrast and the inner working angle. Section 7 is dedicated to results obtained for determining the sensitivity of the coronagraph to the critical components such as the lateral and longitudinal misalignment of the focal plane mask, and the lateral misalignment of the Lyot stop. In Section 8, we briefy discuss the gain in these recent experiments compared to the previous papers presented. Finally, we conclude the paper with a summary of the testbed results, and a description of our future plans toward the development of a full system that will include an adaptive optics for the ground-based observations with large and very large telescopes, and couple it with a spectrograph.

\section{EvWaCo Focal Plane Mask} \label{sec:FPM}
\subsection{The setup}\label{sec:setup}
\begin{figure}[ht!]
\centering
\includegraphics[height=100mm,width=150mm]{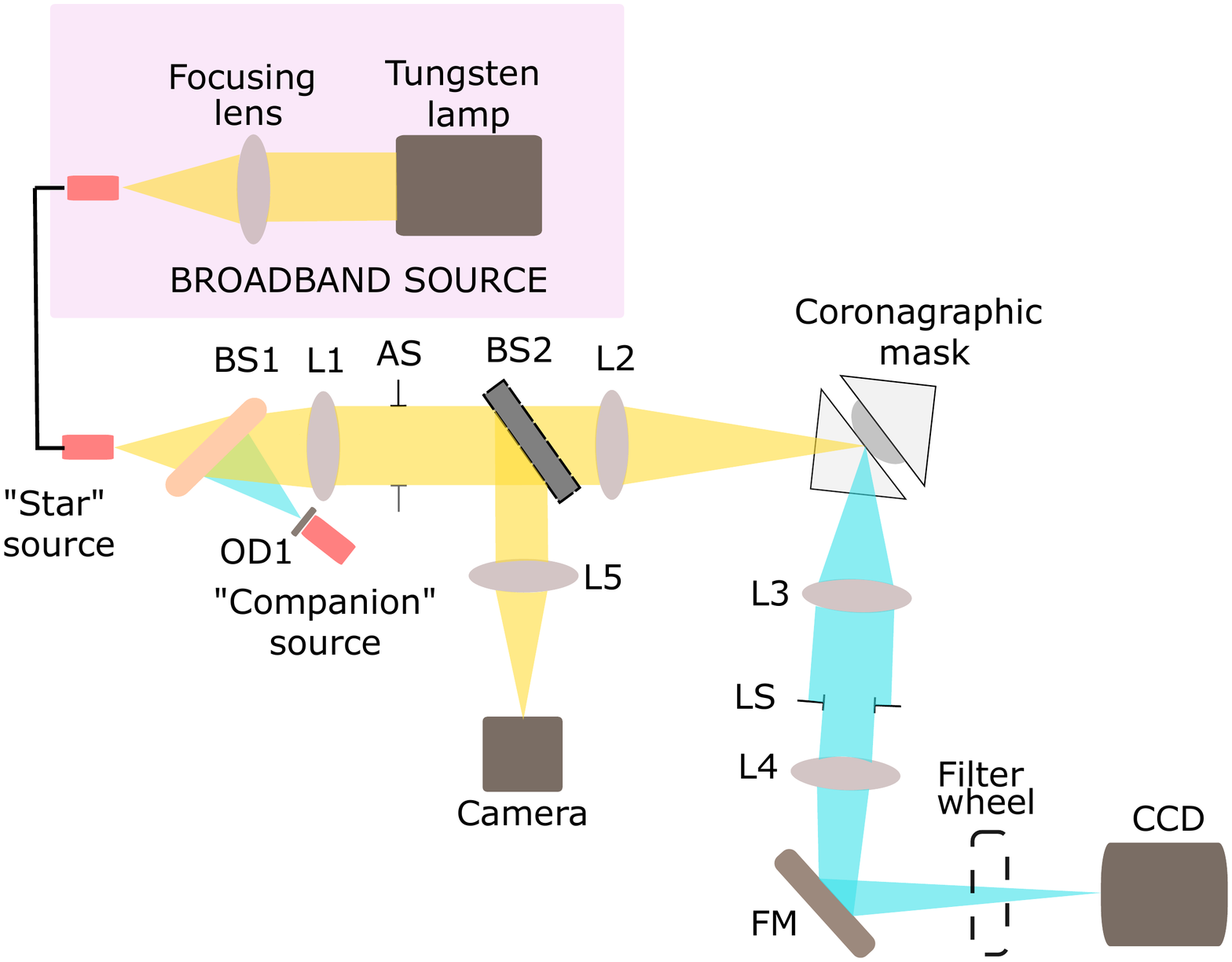}
\caption{Illustration of the EvWaCo setup with the star and companion channels.} \label{fig:EvWaCo_fig1}
\end{figure}

Figure \ref{fig:EvWaCo_fig1} shows an illustration of the current EvWaCo testbed. The setup comprises two sources: the star source and the companion source. The star source is made of a stable tungsten lamp that emits a collimated beam and is focused by an achromatic doublet lens on a 50 $\micron$ multi-mode fiber. This multi-mode fiber is connected to a single mode fiber that directs the light toward the coronagraph bench.

The companion source comprises an LED that emits a beam at the central wavelength $\lambda_c$ = 780 nm with a spectral bandwidth of $\Delta\lambda/\lambda \approx$ 3\%. The light emitted by this LED is injected into a single mode fiber. The beams emitted by the star and the companion fibers are recombined by a beamsplitter BS, collimated by the lens L1 and transmitted through the aperture stop AS of diameter $D_{AS}$. The beam transmitted by the AS is focused by the lens L2 on the Focal Plane Mask (FPM). The on-axis beam incident on the FPM is transmitted while the off-axis beam is reflected by the mask and directed towards the collimating lens L3 and the Lyot Stop LS. The beam transmitted by the LS is focused by the lens L4 and directed by a folding mirror towards the CCD camera. This CCD camera is mounted on a translation stage for fine focusing. A photometric filter, which can either be I-band or R-band, is placed in front of the CCD. 

A filter wheel, placed in front of the camera, houses a neutral density filter OD2 for measuring the unsaturated PSF peak and an achromatic doublet to re-image the pupil plane on the camera. The camera used is an FLI Interline CCD camera cooled at temperature T = -14.6$^{\circ}$ C with a pixel size of 7.4 µm. During the contrast measurements, all the optical components after the lens L2 are enclosed in a black box to protect from stray light.

\begin{figure}[ht!]
\centering
\includegraphics[trim=1cm 4.8cm 1cm 2cm, width=0.95\textwidth, clip]{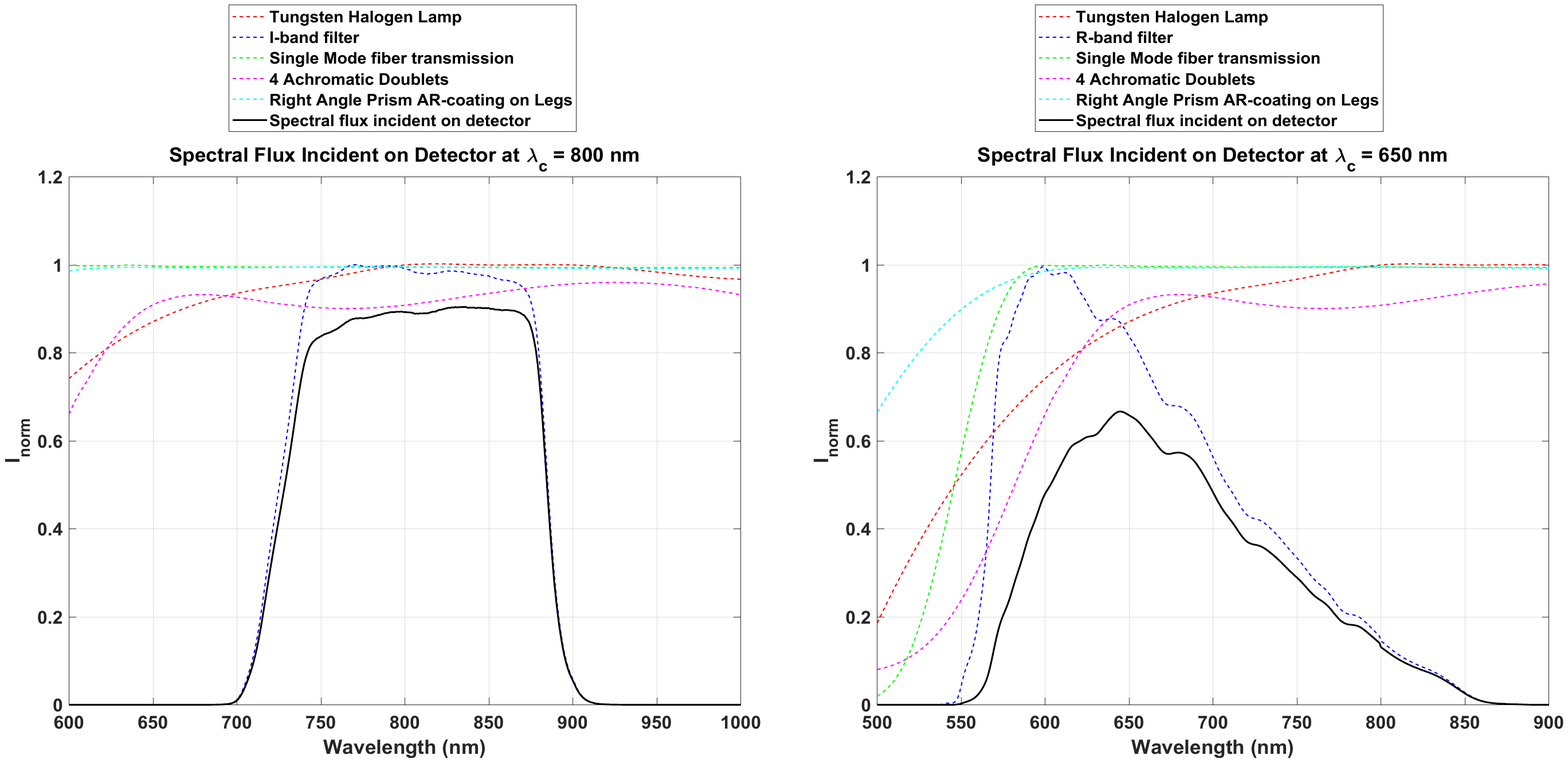}
\caption{Profiles of the spectral flux incident on the the detector at $\lambda_c \approx$ 800 nm (left) and $\lambda_c \approx$ 650 nm (right).} \label{fig:EvWaCo_fig2}
\end{figure}

Figure \ref{fig:EvWaCo_fig2} shows the spectral flux incident on the detector. It takes into account the irradiance due to the tungsten halogen lamp, single mode fiber, transmission coating of the achromatic doublets and the prism used. The relative spectral ratio is approximately equal to 20\% in the I-band and 23\% in the R-band.

\begin{figure}[ht!]
\centering
\includegraphics[trim=0.5cm 0.5cm 0.5cm 0.5cm, width=0.65\textwidth, clip]{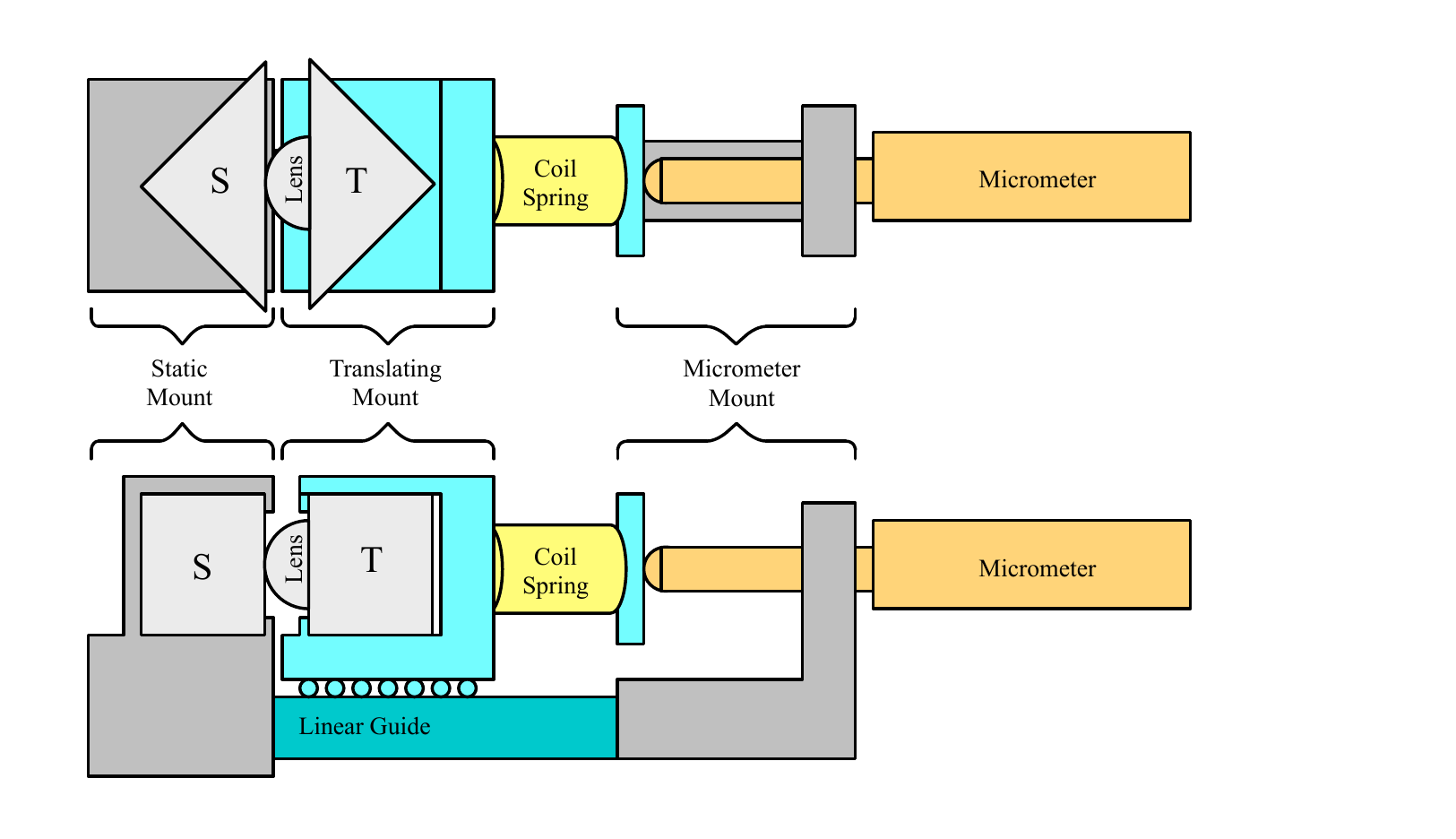}
\caption{Conceptual design of the EvWaCo coronagraphic mask.} \label{fig:EvWaCo_fig3}
\end{figure}

The current EvWaCo testbed has been upgraded with a new coronagraphic mask built according to the design presented in Figure \ref{fig:EvWaCo_fig3}. The static prism S is mounted on the static support while the translating prism is mounted on the carrier moving along the high precision linear slide. The pressure between the lens and the prism is adjusted by turning the micrometer which compresses a coil spring put in contact with the lens. The force over the contact area can be controlled from 0 N to 50 N  with an accuracy equal to 0.02 N. We emphasize that the support of the static prism has been designed in order to keep the line of sight stable while adjusting the pressure between S and T with an accuracy better than 15''. 

\begin{figure}[ht!]
\centering
\includegraphics[trim=1cm 3cm 1cm 1cm, width=0.95\textwidth, clip]{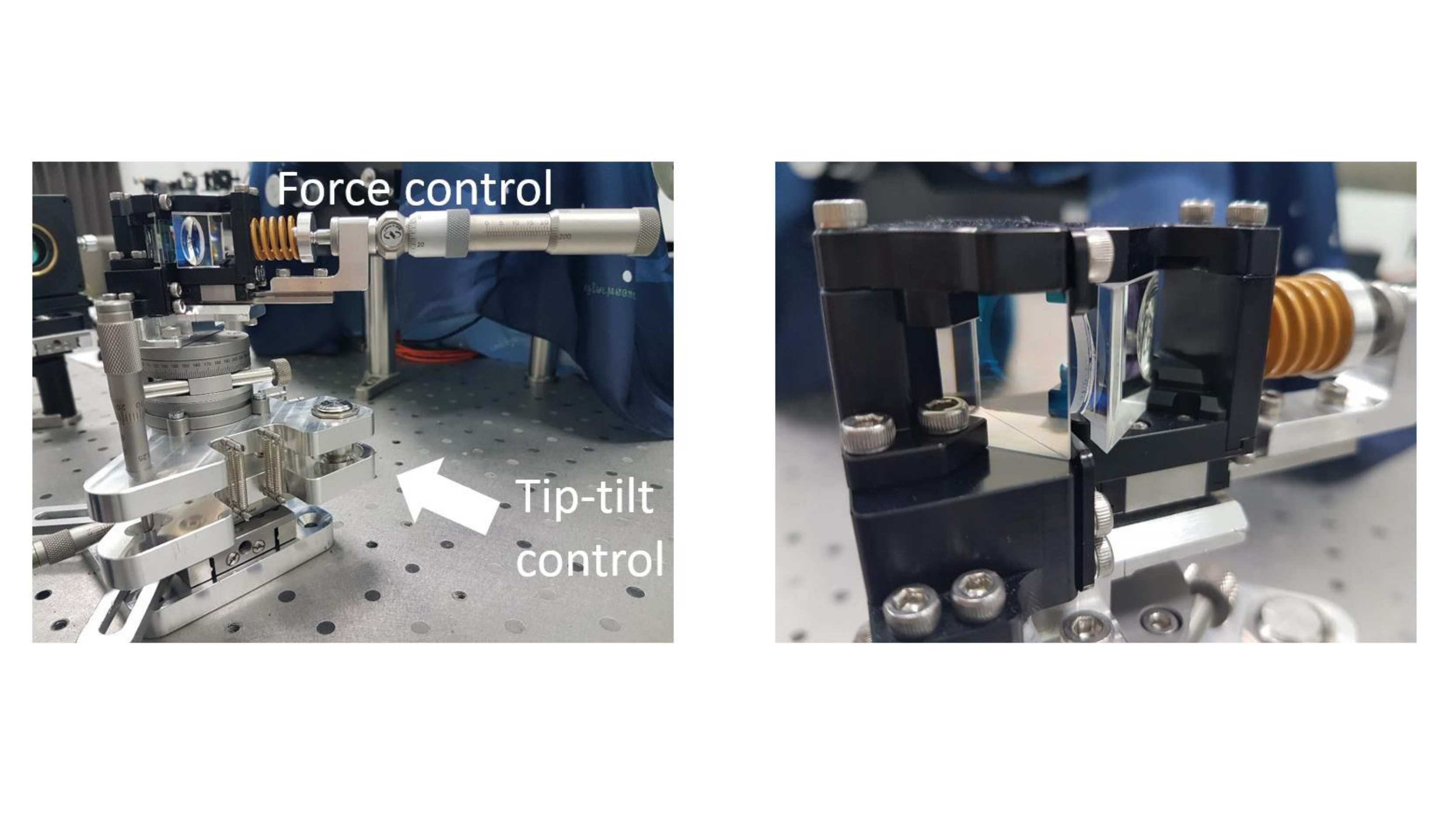}
\caption{Image of the assembled FPM installed on the testbed (left) and a close-up view of the coronagraphic mask (right).} \label{fig:EvWaCo_fig4}
\end{figure}

This coronagraphic mask is mounted on a support (Figure \ref{fig:EvWaCo_fig4}-left) that comprises five degrees of freedom: two translations in the horizontal plane, one rotation around the vertical axis and two rotations around the horizontal axes. The tip-tilt orientation of the mask can be adjusted over a range larger than +/- 5 degrees and with an accuracy better than 10''. This tip-tilt mount is mounted on two translating stages each equipped with a micrometer which is used to position the focal plane mask on the focus with an accuracy equal to 2 µm. Figure \ref{fig:EvWaCo_fig4} (right) shows a close-up view of the coronagraphic mask.

\clearpage

\section{Mask Response Profile and Pupil Irradiance Distribution} \label{sec:mask_reflection}

We measure the flat field irradiance by placing an integrating sphere between the lens L1 and the aperture stop. A narrow-band spectral filter with a central wavelength of $\lambda_c$ is placed between the integrating sphere and the aperture stop. These components are only placed during flat-field measurements; otherwise, they are removed from the setup.

\subsection{Variation of the mask response with the applied pressure}

\begin{figure}[ht!]
\includegraphics[trim=1cm 4cm 1cm 3cm, width=0.95\textwidth, clip]{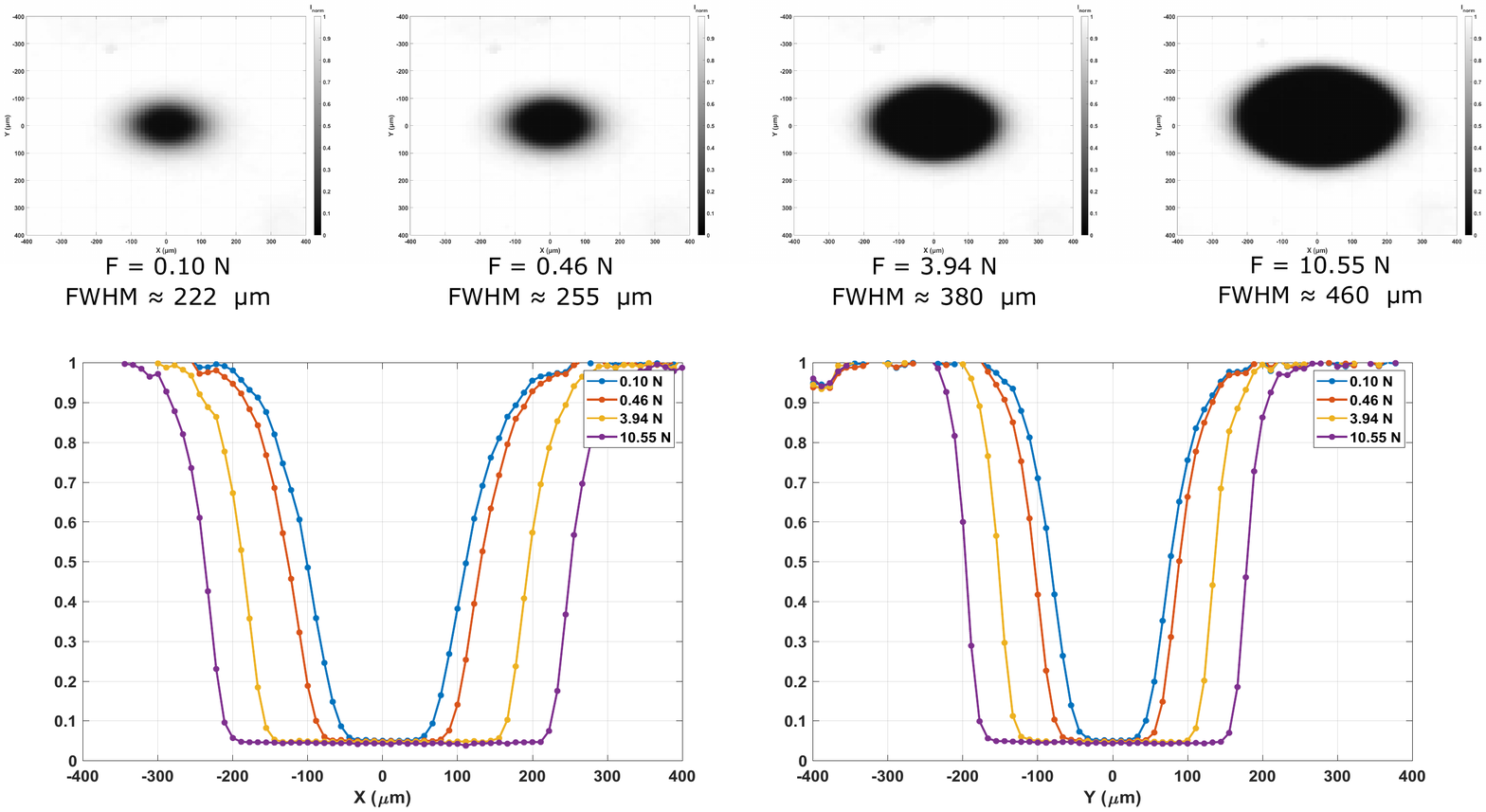}
\caption{[Top]Variation of the shape of the mask response with respect to the increasing force applied on the static prism by the convex lens. These measurements were obtained in unpolarized light using a source of central wavelength equal to $\lambda_c$ = 800 nm and $\Delta\lambda$ = 40 nm. [Bottom] Cross-section along the x-axis (left) and the y-axis (right) of the reflection profile of the mask. As the force applied increases, the size of the mask enlarges, thereby, increasing the IWA and the contact area between the prism and the lens.} \label{fig:EvWaCo_fig5}
\end{figure}

Figure \ref{fig:EvWaCo_fig5} represents the variation of the shape of the coronagraphic mask for a force applying S on P, and varying from 0.10 N and 10.5 N for a source of central wavelength $\lambda_1$ = 800 nm with a Full Width at Half Maximum (FWHM) of 40 nm. At the time of these measurements, the reference is considered to be the case where the lens and the prism are placed in contact without applying any force. 

There is no force sensor in the setup so the force is calculated using the configuration of a sphere place in contact with a flat plate (Hertzian contact stress) \citep{Shigley}. We calculate the contact stress (diameter of the hole) that corresponds to a specific contact pressure represented by the micrometer readings. This results to a force profile given the actuator reading, and thus, relate the diameter of the contact area to the force/pressure applied. Figure \ref{fig:EvWaCo_fig5} also shows the x- and y- profiles of the mask response to increasing pressure. It can be observed that as pressure on the prism increases, the size of mask increases in terms of the FWHM and the contact area.

In this paper, the inner working angle (IWA) is defined as the distance from the star center where the post-coronagraphic throughput is 50\%. Thus, it is expressed in terms of $\lambda/D$ where D is the diameter of the aperture stop. Since the size of the mask directly impacts the IWA, by varying the force applied by the lens on the prism, the IWA can be finely tuned as well. For example, an FWHM of 222 µm corresponds to 3.7 $\lambda/D$ at $\lambda_c \approx$ 800 nm.

The pressure of the lens has been adjusted until an IWA approximately equal to $3 \lambda_1/D$ is achieved along the minor axis for a source central wavelength equal to $\lambda_1 \approx$ 800 nm. All the experiments presented hereafter have been performed under this pressure unless specified otherwise.

\clearpage

\subsection{Modelling the coronagraphic mask}
To model this coronagraphic mask, the air thickness, d, is approximated by a polynomial equation given by Equation \ref{eqn:air_gap}:

\begin{equation}
d(\rho) = max(a_{1}\rho + a_{2}\rho^{2},0),
\label{eqn:air_gap}    
\end{equation}

where $a_1$ and $a_2$ are the coefficients of the polynomial of degrees 1 and 2, respectively, and $\rho$ is the radial distance to the contact point in the prism hypotenuse plane. We calculate the theoeretical reflection coefficients by applying the coefficients locally in the case of thin, homogenous air gap with an optical index, $n_2$, sandwiched between two optical indices, $n_1$, and $n_3$ \citep{Buisset, Zhu}. The coefficients of the polynomial are calculated by minimizing the sum of the squared difference between the simulation and the experimental data. We use a built-in function in Matlab called \textit{fminsearch} that uses the Nelder-Mead simplex direct search \citep{Lagarias} to perform the optimization. Using this optimization, the coefficients of the polynomial are found to be $a_1 = -1.10^{-3}$ and $a_2$ = 36 $\micron^{-1}$. 

\begin{figure}[ht!]
\centering
\includegraphics[trim=0.5cm 9.8cm 0.5cm 4cm, width=0.95\textwidth, clip]{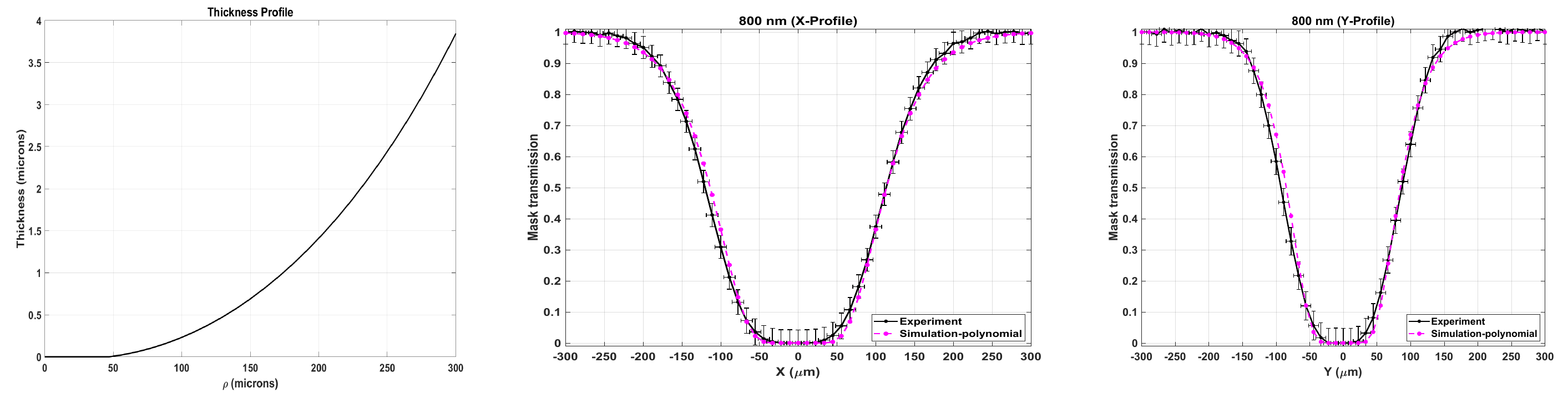}
\caption{[Left] Approximation of the air gap at the contact area between the lens and the prism. [Middle] Comparison between the theoretical and measured mask response along the major axis of the mask (x-axis). [Right] Comparison between the measured mask response  along the minor axis (y-axis). The error bars along the horizontal axis is equal to $\pm 1$ pixel while the error bars along the vertical axis is the standard deviation of the irradiance value from the 101 flat-field images obtained at the time of the acquisition of the experimental data.} \label{fig:EvWaCo_fig6} 
\end{figure}

Figure \ref{fig:EvWaCo_fig6} compares the theoretical and the experimental mask reflectivity for unpolarized light.  The extracted profiles from the experimental data are normalized with respect to the median value of the calibrated image that is measured at a distance equal to 400 µm (13 $\lambda_1/D$) from the optical axis of the coronagraphic mask. The horizontal error bars represent $\pm$ 1 pixel while the vertical error bars represent the standard deviation $\pm \sigma$ of 101 intensity measurements.

The mask profiles show that the FWHMs along X and Y are $FWHM_x$ $\approx$ 240 µm and $FWHM_y$ $\approx$ 180 µm, respectively. Consequently, the measured IWAs are as follows: $IWA_x$ $\approx$ 4.00 $\lambda_1/D$ and $IWA_y$ $\approx$ 3.00 $\lambda_1/D$ at the wavelength $\lambda_c$ = 800 nm.

This enlargement of the mask along the X axis is due to the angle of incidence on the coronagraphic mask prism hypotenuse as discussed in our previous paper (Buisset et al. 2017). We verified that the ratio $IWA_{x}/IWA_{y} = 1/cos(\theta)$  is equal to  1.33 that is compatible with an angle of incidence approximately equal to $\theta = 41.4^{\circ}$.

At distances comprised between -50 µm and +50 µm along the x-axis and between -30 µm and +30 µm along the y-axis, we notice that the theoretical mask transmission is slightly smaller than the experimental one. This difference is close to the error bar but could explain the discrepancy between the theoretical and the experimental raw contrasts at distances smaller than 3 $\lambda/D$ as presented in Section 5 and in Figure 10. 

At distances comprised between +50 µm and +150 µm along the x-axis and between +70 µm and +125 µm  along the y-axis, we notice a good match between the theoretical model and the experimental results, and the FWHM is correctly modelled. At distances greater than 150 µm in the x-axis and 120 µm in the y-axis, we notice a slight discrepancy between the theory and the experimentation. However, at these distances, we expect that this small difference will only have a small impact on the raw contrast calculation. 

\clearpage
\subsection{Mask response at different wavelengths}

We, then, investigate the response of the new focal plane mask to a different wavelength, $\lambda_2$ = 650 nm. To measure the mask response at $\lambda_2$, we used an LED emitting at $\lambda_{peak}$ = 617 nm and a spectral filter centered at $\lambda_2$ = 650 nm with an FWHM equal to 40 nm. 

\begin{figure}[ht!]
\centering
\includegraphics[trim=1cm 1cm 2.5cm 1cm, width=0.85\textwidth, clip]{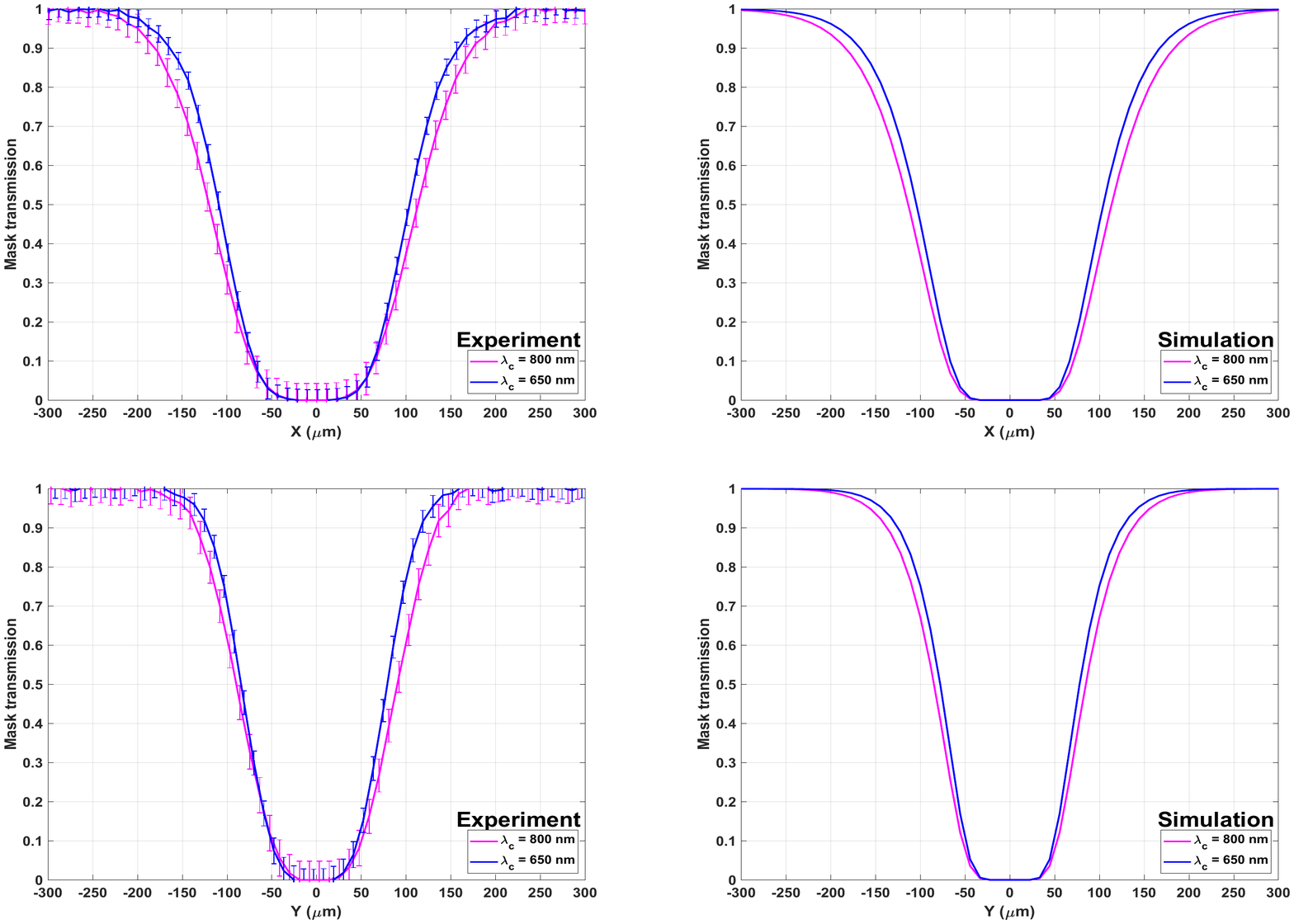}
\caption{(First column) Experimental mask profiles along the x- and the y-axis at different central wavelengths wavelengths:  $\lambda_1$ = 800 nm (pink curve) and  $\lambda_2$ = 650 nm (blue curve). (Second column) Theoretical mask profiles at these same wavelengths.} \label{fig:EvWaCo_fig7}
\end{figure}

Figure \ref{fig:EvWaCo_fig7} shows the theoretical and experimental mask response profiles of the focal plane mask obtained using the spectral filter with central wavelength, $\lambda_2$, and compared it to the mask response at $\lambda_1$. The theoretical mask response is calculated by using the same polynomial coefficients obtained at 800 nm.  We observe that the mask profiles are smaller at $\lambda_2$ (blue curve)  compared to the ones obtained at $\lambda_1$ (pink curve). In terms of the IWA, the values are quite close. At $\lambda_2$,  $IWA_X \approx$ $4.27$ $\lambda/D$ ($FWHM_X \approx$ $208$ \textmu m) and $IWA_Y \approx$ $3.24$ $\lambda/D$ ($FWHM_Y \approx$ $158$ \textmu m). 

These results show that the mask itself adapts well to the wavelengths between $\lambda_1$ and $\lambda_2$ that correspond to the central wavelengths of the I- and R- bands of the Johnson-Cousins photometric filter. These first experimental observations of the quasi-achromatic behavior of the mask confirms the theoretical prediction for two wavelengths separated by 150 nm. 

\subsection{Pupil Irradiance Distribution}

Figure 8 represents the irradiance distribution at the exit pupil plane when the PSF of the star source is placed off-axis (left) and when it is centered on the mask (right). These images were obtained by placing a lens, L5, inside the filter wheel (see Figure 1) in order to image the exit pupil. The LS diameter is equal to 81\% of the exit pupil diameter and transmits approximately 65\% of the incident light. The results show that the energy is mostly concentrated at the edges, forming a fine annular shape with a diameter that is approximately 2 mm, and an average extension width equal to 86 µm.

\begin{figure}[ht!]
\centering
\includegraphics[trim=1cm 4cm 1cm 1cm, width=0.65\textwidth, clip]{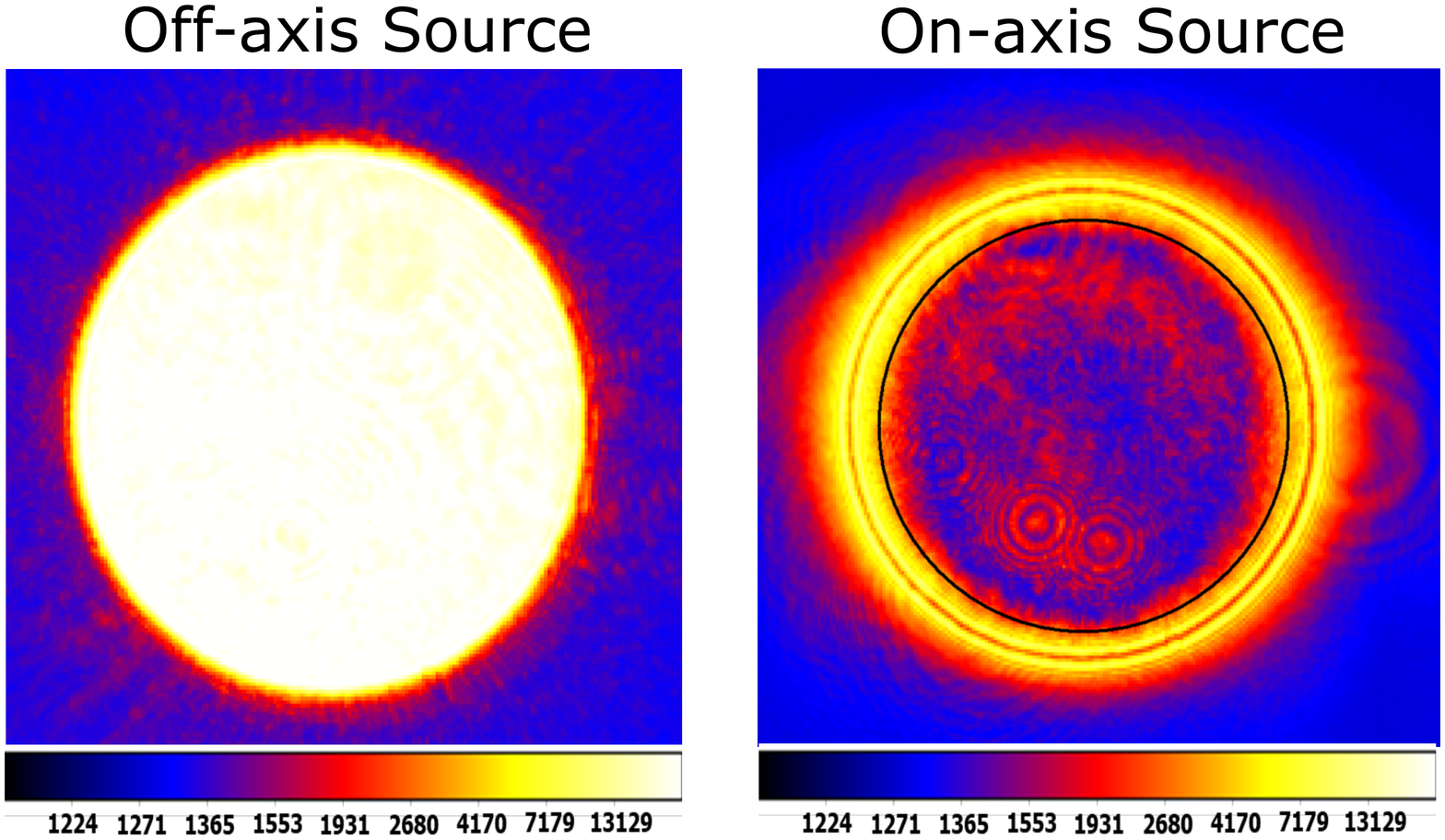}
\caption{Irradiance distribution at the exit pupil when the PSF is placed off-axis (left) and when it is centered on the mask (right). The images were obtained using a source of central wavelength $\lambda_C$ = 780 nm and a spectral ratio of $\Delta\lambda/\lambda \approx$ 3\%. The black circle represents the position of the Lyot stop. The Lyot stop diameter is approximately 81\% the pupil diameter.} \label{fig:EvWaCo_fig8}
\end{figure}

\clearpage

\section{Deep Contrast Measurements}
\subsection{Raw contrast performance}

\begin{figure}[ht!]
\centering
\includegraphics[trim=0.2cm 8cm 0.5cm 3cm, width=0.95\textwidth, clip]{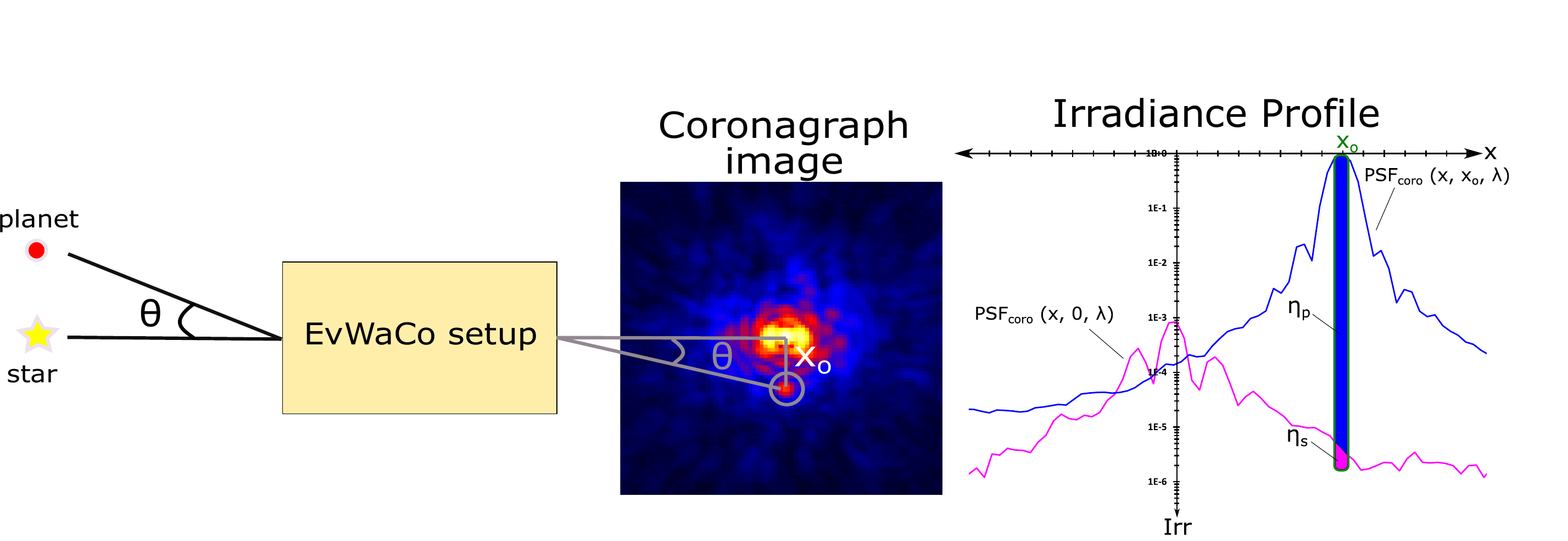}
\caption{Qualitative schematic for calculating the raw contrast. The star and the planet are separated by an angle, $\theta$. The coronagraph image shows the star centered on the coronagraphic mask with together with a planet located at a distance, $x_o$, from the star center.} \label{fig:EvWaCo_fig9}
\end{figure}

Figure \ref{fig:EvWaCo_fig9} illustrates the calculation of contrast adapted in this paper. We decided to take the definition of the raw contrast presented in \citep{Ruane}. Thus, the raw contrast is calculated as:

\begin{equation}
C(x,\lambda) = \frac{\eta_s(x_o,\lambda)}{\eta_p(x_o,\lambda)},
\end{equation} 

where $\eta_s(x_o,\lambda)$ is the fraction of available starlight detected and $\eta_p(x_o,\lambda)$ is the fraction of companion light detected. The quantity $\eta_s(x_o,\lambda)$ is given by 

\begin{equation}
    \eta_s(x_o,\lambda) = \int_{AP(x_o)} PSF_{coro}(x,0,\lambda) dx,
\end{equation}

where $PSF_{coro} (x,0,\lambda)$ is the point spread function (PSF) centered on the focal plane mask recorded at the image plane and AP($x_o$) is equal to 1 pixel. Meanwhile, the quantity $\eta_p(x_o,\lambda)$ is calculated as

\begin{equation}
    \eta_p(x_o,\lambda) = \int_{AP(x_o)} PSF_{coro}(x,x_o,\lambda) dx. 
\end{equation}

where $PSF_{coro}(x,x_o,\lambda)$ is the PSF recorded at a distance $x_o$ from the center of the mask. We assume that $PSF_{coro}$ is invariant over the FOV of the coronagraph at distances higher than 7 $\lambda/D$ where the mask transmission is close to 1.

In Figure \ref{fig:EvWaCo_fig9} (left), we show an image of the star residual irradiance $PSF_{coro}(x,0, \lambda)$ and the PSF of the companion $PSF_{coro}(x,x_o, \lambda)$ measured at a distance $x_o =$ 8 $\lambda/D$ from the mask center. On the right, we illustrate the calculation of $\eta_s$ and $\eta_p$ from the cross-section of the $PSF_{coro}(x,0, \lambda)$ and $PSF_{coro}(x,x_o, \lambda)$, respectively.

In our case, the fraction of the available starlight is the normalized irradiance, $I_{norm}$, that is the ratio between the unsaturated off-axis PSF peak and the on-axis PSF. Discussion on the calculation of $I_{norm}$ from the experimental data can be found in the Appendix. The light detected from the companion, on the other hand, depends on the transmission, $T(x,\lambda)$, of the mask. The raw contrast is, then, calculated as 

\begin{equation}
C(x,\lambda) = \frac{I_{norm(x,\lambda)}}{T(x,\lambda)}.
\end{equation} 

\begin{figure}[ht!]
\centering
\includegraphics[trim= 0.5cm 2.8cm 0.5cm 1cm, width=0.95\textwidth, clip]{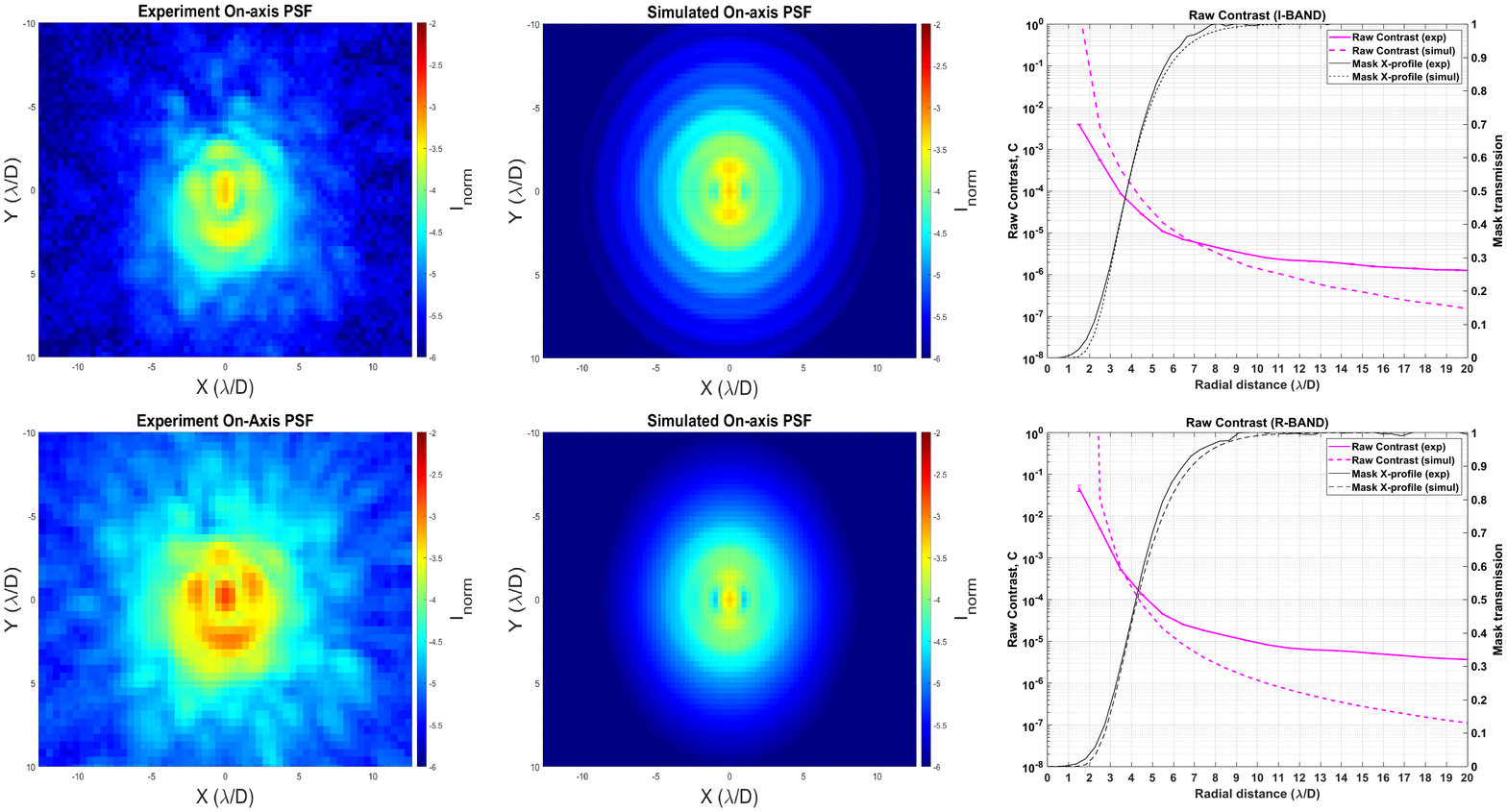}
\caption{Experimental (Column 1) and simulated (Column 2) normalized irradiance distribution of the on-axis star over the full I-band (Row 1) and R-band (Row 2). The third column shows the calculated raw contrast for each passband obtained both experimentally and numerically.} \label{fig:EvWaCo_fig10}
\end{figure}

We show in Figure \ref{fig:EvWaCo_fig10} the normalized irradiance and raw contrast obtained using two spectral bands of the Johnson-Cousins photometric filter \citep{Bessel}: I-band and the R-band. The I-band filter has a central wavelength $\lambda_c$ = 800 nm with a wavelength range between 700 nm and 900 nm and a spectral ratio $\Delta\lambda/\lambda_C$ = 20\%. Meanwhile, the R-band filter has a central wavelength of $\lambda_c$ = 650 nm with a wavelength range between 550 nm and 800 nm and a spectral ratio $\Delta\lambda/\lambda_C$ = 23\%. We note that only the bandpass filter located in front of the camera during the acquisitions over the full I-band or the R-band is changed.

We superimpose on the same figure the results of the simulations done using the same model described in \citep{Buisset1} represented by the dashed lines with an updated mask transmission. This model takes the following inputs: i) spectral flux incident on the detector (see Figure 2), ii) the air thickness profile (see Figure 6), iii) the diameter of the AS and LS, and iv) the focal length of L2, L3 and L4. The complex amplitude incident on the FPM is obtained by taking the Fourier transform of the complex amplitude transmitted by the AS while the complex amplitude incident on the CCD is obtained by taking the Fourier transform transmitted by the LS. A full description of this model can be found in \citep{Buisset1}.

Figure \ref{fig:EvWaCo_fig10} shows the radial average of the raw contrast measured at radial distances between 1.5 $\lambda/D$ to 20 $\lambda/D$ (pink curve). We superimpose in the same graph the mask response curves obtained through the experiment and simulations. For both spectral bands, the coronagraph reduce the PSF peak by three orders of magnitude.

Between 3 $\lambda/D$ and 4 $\lambda/D$, the raw contrast in the I-band varies between $ 2.10^{-4}$ and  $ 5.10^{-5}$ while it varies between $ 2.10^{-3} $ and $ 3.10^{-4}$ in the R-band. Meanwhile, along the same distance, the theoretical raw contrast varies between $ 1.10^{-3}$ and $ 2.10^{-4}$ in the I-band and between $ 3.10^{-3}$  and $ 2.10^{-4}$ in the R-band. Between 4 $\lambda/D$ and 5 $\lambda/D$, the raw contrast in the I-band varies between $ 5.10^{-5}$ and  $ 2.10^{-5}$ while it varies between $ 3.10^{-4} $ and $ 8.10^{-5}$ in the R-band. On the other hand, the theoretical raw contrast at the same distance varies between $ 1.10^{-4}$ and $ 3.10^{-5}$ in the I-band and between $ 2.10^{-4}$  and $ 4.10^{-5}$ in the R-band. We, thus, notice a difference between the theoretical model and the experimental data. The potential origin of this discrepancy could be to the mismatch between the theoretical and the experimental mask transmission as discussed in the previous section. Although the experimental raw contrast values between the two bandpass differs by an order of a magnitude, the theoretical results show that the raw contrast performance is similar in the R- and I-bands, thus, confirming the theoretical achromaticity of EvWaCo. We suspect that the presence of low-order aberrations close to the IWA contributed to the discrepancy in the experimental raw contrast value.

Between 5  $\lambda/D$ and 10 $\lambda/D$, the contrast decreases from C $\approx$ $2.10^{-5}$ to C $\approx$ $3.10^{-6}$ for the I-band and from C $\approx$ $8.10^{-5}$ to C $\approx$ $9.10^{-6}$ for the R-band. Along the same distance, the theoretical contrast is between C $\approx$ $10^{-5}$ to C $\approx$ $10^{-6}$ for both the I-band and the R-band. We believe that the origin of this difference between the experiment and the theoretical contrast is the presence of speckles in the experimental measurements. Overall, our theoretical results show that the similarity between the I- and R-bands is valid at every radial distance. This confirms one more time the theoretical achromatic behavior of the EvWaCo mask in unpolarized light over the two spectral bands.

\subsection{Raw contrast at different IWA}

\begin{figure}[ht!]
\centering
\includegraphics[trim= 1cm 0.3cm 1cm 0cm, width=0.95\textwidth, clip]{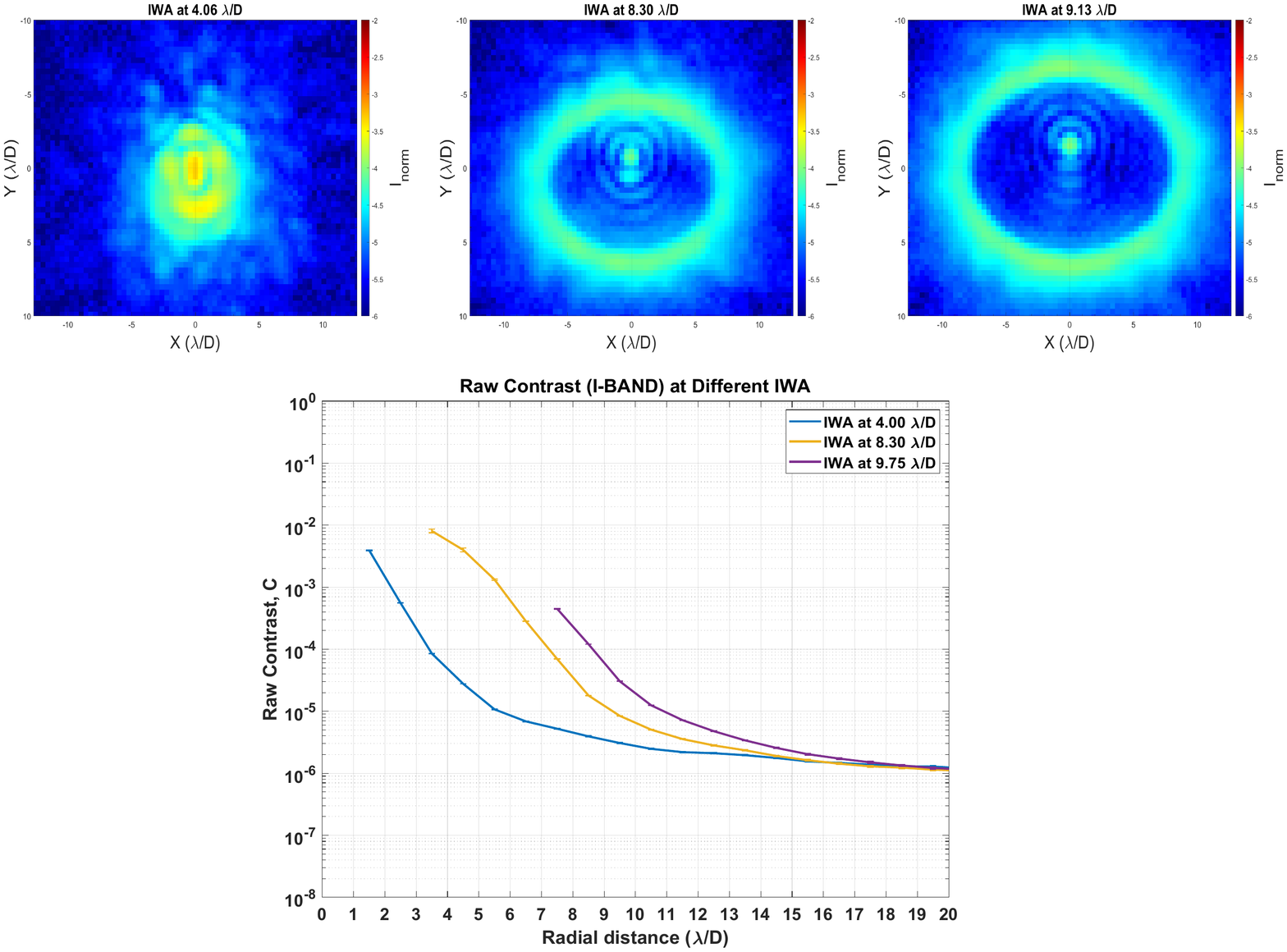}
\caption{[Top] 2D map of the normalized irradiance distribution of the on-axis star over the full I-band at different IWA. [Bottom] Radial average of the raw contrast curves. Refer to Figure 5 for the mask response at each IWA.} \label{fig:EvWaCo_fig11}
\end{figure}

Figure \ref{fig:EvWaCo_fig11} shows the experimental raw contrast obtained at different IWA. We notice that the best raw contrast performance is always reached with the FPM with the smallest IWA.
It is important to mention that : 

i) The FPM transmission profile is quasi-gaussian only when the contact area is small and when the IWA is close to 3 $\lambda/D$ (minor axis). This quasi-gaussian shape has the reputation to provide a better contrast performance than the sharp-edge transmissions obtained at larger pressure (Figure 5), and 

ii) The achromatic property of this coronagraph is maximum at smaller IWA. Indeed, the chromatic variation of the mask response is valid only at the distance where the mask transmission varies between 0 and 1.
Hence, we deduce that masks with a small IWA will provide a quasi-gaussian response that fully adapt itself to the wavelength. The masks with a large IWA will provide a top-hat transmission that does not change with the wavelength.

During the astronomical observations, the size of the proposed FPM could thus be adjusted in order to provide:

i) A small IWA close to 3 $\lambda/D$ to reach the  best contrast performance during excellent seeing conditions of star environment imaging and companion detection and characterization, and

ii) A large IWA comprised between 3 $\lambda/D$ and 8 $\lambda/D$ during bad seeing conditions to improve the low order wavefront distortion measurement by a WFS installed  on the star channel \citep{Buisset1} and to better correct the contrast degradation induced by the turbulence.

The large IWA could also be used to suppress the halo of extended objects with an acceptable loss of contrast performance, for example, in Cosmic  Origins Science observations as mentioned in \cite{Ebbets}.

\clearpage
\subsection{ DETECTION OF A COMPANION}\label{sec:detec_faint}

We consider that the coronagraph is used to detect a faint companion with its PSF located in the vicinity of the star PSF that is itself set on-axis. The detection of the companion, thus, consists of extracting the companion signal from the background noise induced by the following contributors: star residual photon noise, speckle noise, stray light photon noise, detector thermal noise and readout noise. We assess the level of this background noise by calculating the standard deviation, $\sigma_{Background}$, over an annular region starting with an inner radius R1 = 2.5 $\lambda$/D and an outer radius,  R2 = 3.5 $\lambda$/D. This calculation is repeated over an annular region with inner radii equal to 3.5, 4.5,…, 19.5 $\lambda$/D,  and with an annular width that is always kept equal to 1 $\lambda$/D.

\begin{figure}[p]
\centering
\includegraphics[trim=0.1cm 6cm 0.1cm 4cm, width=0.95\textwidth, clip]{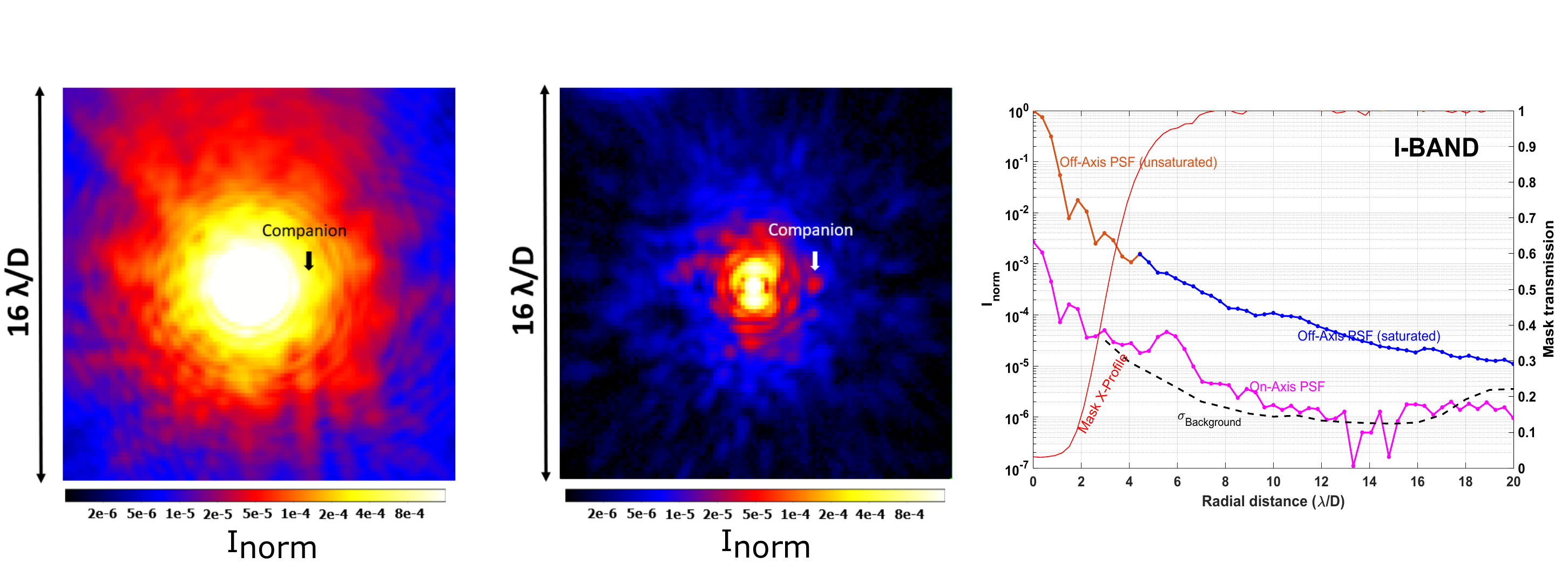}
\caption{Irradiance distribution of the star with a companion, 25000 times fainter than the star, located at 5.5 $\lambda/D$ from the star center, without the coronagraph (first figure), with the coronagraph (second figure), and cross-section along the x-axis of the irradiance distribution of the star and the companion (third figure). In this experiment, the star source is in the I-band and the companion source is an LED diode emitting at $\lambda_C$ = 780 nm with a spectral ratio $\Delta\lambda/\lambda \approx$ 3.2\%. The images are a median of 101 calibrated science image obtained at an integration time of 1 s per frame. The SNR calculated at the companion location is equal to 11, illustrating the detectability of the companion.} \label{fig:EvWaCo_fig12}
\end{figure}

On Figure \ref{fig:EvWaCo_fig12}, we have represented the level of the background noise at distances comprised between 3 $\lambda$/D and 20 $\lambda$/D (black dashed curve). Between 3 $\lambda$/D and 10 $\lambda$/D from the center of the mask, the background noise varies from $\sigma_{Background}$ $\approx$ $3.10^{-5}$ to $\sigma_{Background}$ $\approx$ $10^{-6}$. At distances greater than 10 $\lambda$/D, this noise decreases from $\sigma_{Background}$ $\approx$ $10^{-6}$ to $\sigma_{Background}$ $\approx$ $10^{-7}$. The background noise comprises the detector noise, the star residual photon noise, the stray light photon noise and the speckle noise.

We place the PSF of a companion, 25,000 times fainter than the star, at a location 5.5 $\lambda$/D distance from the center of the star. Figure \ref{fig:EvWaCo_fig12} also shows the irradiance distribution for both the star and the companion with and without the coronagraphic effect. Obviously, when the star is off-axis, the companion is barely detected in the blinding light of the star. However, once the star is centered on the coronagraphic mask, the companion becomes visible.

The detectability of the companion is verified by calculating the signal-to-noise ratio at the location of the companion. To calculate the background noise, $\sigma_{Background}$, at the location of the companion $x_o$, we take the standard deviation of a circle with a radius equal to 1 $\lambda$/D, centered at $x_o$. Figure \ref{fig:EvWaCo_fig12} shows the residual light profiles without (left panel) and with (right panel) the companion. We estimate that at the location of the companion, the SNR is equal to 11.

\section{Repeatability and Stability of the Measurements}\label{sec:repeatability}

We verified the stability of the FPM and the EvWaCo setup, and the repeatability of the reported contrast by measuring the performance of the coronagraph over 8 months. In each measurement, the PSF was re-centered on the FPM. The pressure between the lens and the prism was the same throughout the 8 months. However, the lateral position of the Lyot stop was readjusted from time to time so that it remains centered on the exit pupil. 

\begin{figure}[p]
\centering
\includegraphics[trim=1cm 6cm 1cm 5cm, width=0.95\textwidth, clip]{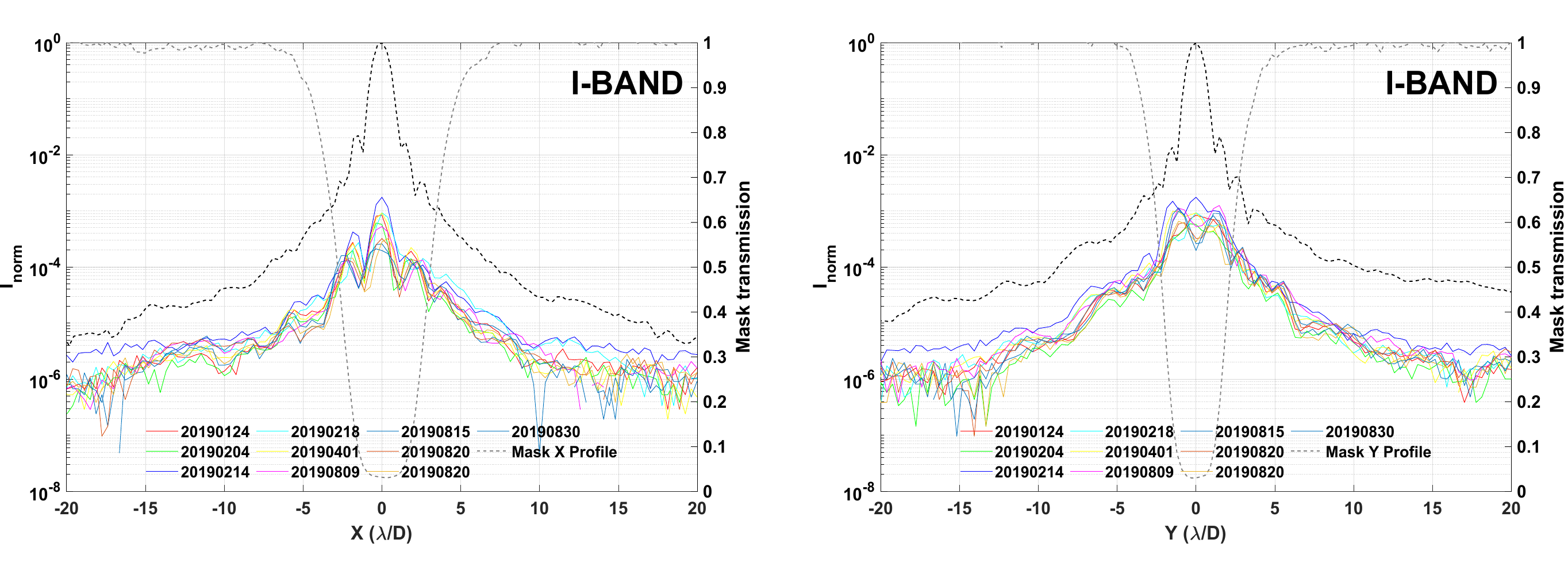}
\caption{Cross-section along the x-axis (left) and the y-axis (right) of the PSF images obtained over the full I-band in unpolarized light. The different colors represent 10 normalized irradiance measurements done over 8 months. It can be observed that the overall shape of the X- and Y-profiles is retained over this time period.} \label{fig:EvWaCo_fig13}
\end{figure}

\begin{figure}[p]
\centering
\includegraphics[trim=2cm 2cm 1.5cm 0.5cm, width=0.55\textwidth, clip]{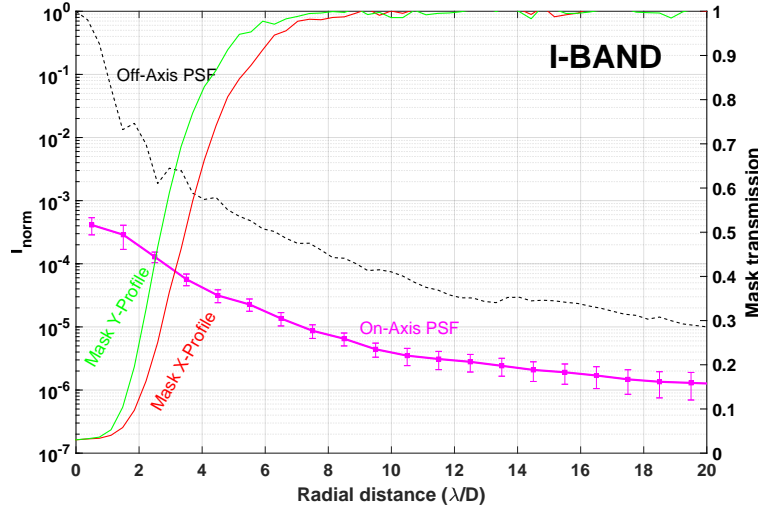}
\caption{Radial average of the normalized irradiance and the corresponding standard deviation of the 10 contrast measurements. At the location of the IWA, the average normalized irradiance is approximately equal to $I_{norm} \approx 6.10^{-5}\pm 1.10^{-5}$ at 3.5 $\lambda/D$, $I_{norm} \approx 3.10^{-5} \pm 7.10^{-6}$ at 4.5 $\lambda/D$ and $I_{norm} \approx 2.10^{-5} \pm 5.10^{-6}$ at 5.5 $\lambda/D$.} \label{fig:EvWaCo_fig14}
\end{figure}

Figure \ref{fig:EvWaCo_fig13} shows the x- and y- profiles of the residual light pertaining to the I-band. The black curve represents the average of these measurements. We observe that the overall shape of the profile remains the same despite taking the measurements at different dates. Figure \ref{fig:EvWaCo_fig14} shows the radial average of the curves for 10 sets of measurements. At the location of the shortest IWA, the average normalized irradiance is $I_{norm} \approx 10^{-4}$.The average normalized irradiance is equal to $I_{norm} \approx 6.10^{-5} \pm 1.10^{-5}$ at 3.5 $\lambda/D$, $I_{norm} \approx 3.10^{-5} \pm 7.10^{-6}$ at 4.5 $\lambda/D$ and $I_{norm} \approx 2.10^{-5} \pm 5.10^{-6}$ at 5.5 $\lambda/D$. Note that these measurements were obtained on a passive setup without adaptive optics.

\clearpage
\section{Sensitivity Measurements}
\subsection{Variation of the normalized irradiance due to tip-tilt}

When observing the environment of stars with large telescopes equipped with AO, it is critical that the PSF is properly centered on the focal plane mask; otherwise, it will contribute to the contamination of the faint companion signal \citep{Lloyd}. The measurement of the variation of the contrast performance is done by installing a pointing camera. A beamsplitter is installed right after the aperture stop and the beam reflected is focused by the lens L5 on a Point Grey Camera with a pixel size of 3.4 $\micron$ as shown in Figure \ref{fig:EvWaCo_fig1}. 

\begin{figure}[ht!]
\centering
\includegraphics[trim=1cm 0.3cm 1cm 0cm, width=0.95\textwidth, clip]{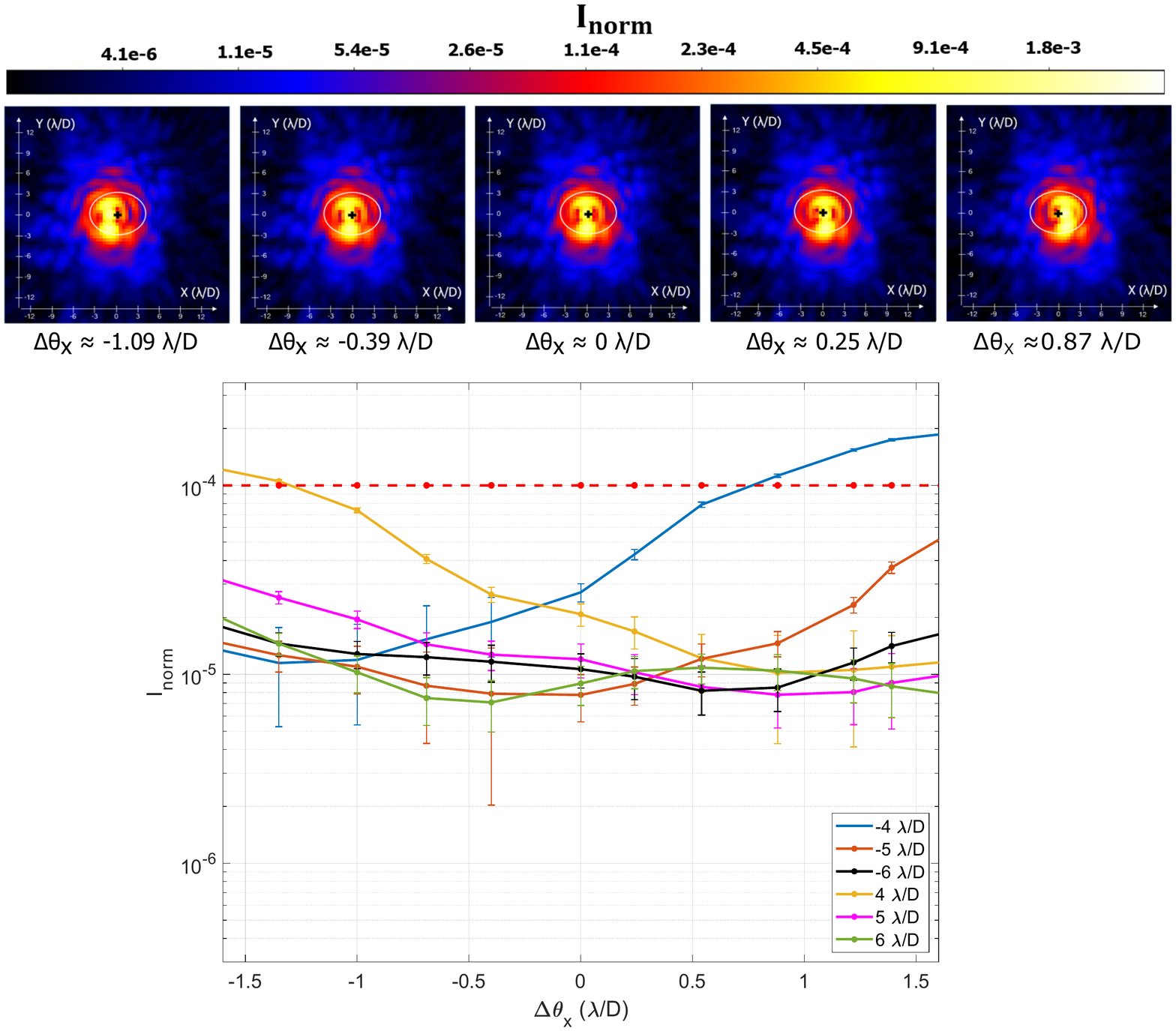}
\caption{[Top panel] Evolution of the irradiance distribution recorded by the CCD camera at the companion channel as tilt along the x-axis is introduced in the testbed. The black cross indicates the center of the mask and the white ellipse represents the location of the IWA. The images are shown at the same visualization level. [Bottom panel] Cross-section along the x-axis of the normalized irradiance taken at the following locations: $\pm 4 \lambda/D$, $\pm 5 \lambda/D$ and $\pm 6 \lambda/D$. The error bars correspond the standard deviation of the normalized irradiance from the 101 images acquired for each tilt.} \label{fig:EvWaCo_fig15}
\end{figure}

\begin{figure}[ht!]
\centering
\includegraphics[trim=1cm 0.2cm 1cm 0cm, width=0.95\textwidth, clip]{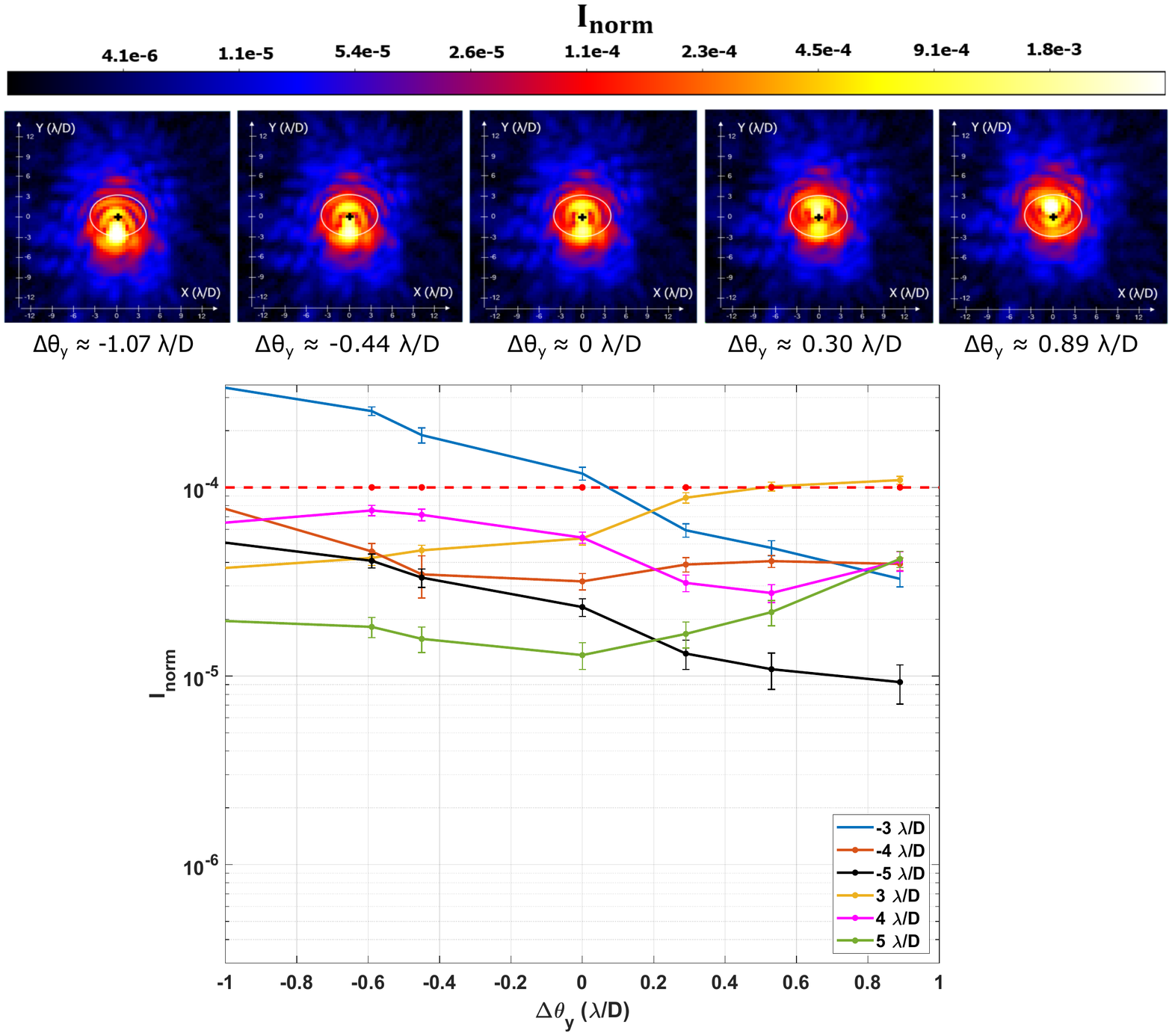}
\caption{[Top panel] Evolution of the irradiance distribution recorded by the CCD camera at the companion channel as tilt along the y-axis is introduced in the testbed. The black cross indicates the center of the mask and the white ellipse represents the location of the IWA. The images are shown at the same visualization level. [Bottom panel] Cross-section along the x-axis of the normalized irradiance taken at the following locations: $\pm 3 \lambda/D$, $\pm 4 \lambda/D$ and $\pm 5 \lambda/D$. The error bars corresponds to the standard deviation of the normalized irradiance from the 101 images acquired for each tilt.} \label{fig:EvWaCo_fig16}
\end{figure}

First, the PSF is centered on the mask which serves as the reference image. Then, tilt is introduced by de-centering the PSF on the FPM. This is done by controlling the lateral alignment of the star source fiber output face. We show in Figures \ref{fig:EvWaCo_fig15} and \ref{fig:EvWaCo_fig16} the irradiance distribution and the respective x- and y- profiles recorded by the CCD at the companion channel. The tilt is expressed in terms $\lambda/D$ where D is equal to the diameter of the aperture stop The error bar corresponds to the standard deviation of the contrast values calculated from the 101 images obtained for each tilt.

Along the x-axis, we notice that when the tilt is smaller than $\theta_X < 1$ $\lambda/D$, the normalized irradiance at the IWA is better than $10^{-4}$. Meanwhile, both the normalized irradiance at 5 $\lambda$/D and 6 $\lambda$/D show slight variations. At $\pm 3$ $\lambda/D$, the cross-section along the y-axis, on the other hand, shows an asymmetry in the normalized irradiance profile. This is due to the fact that the PSF position set as the reference is slightly shifted along the y-axis. Despite that, we can deduce from Figure \ref{fig:EvWaCo_fig16} that normalized irradiance is better than $10^{-4}$ for tilt angles smaller than $\theta_Y < 0.2$ $\lambda/D$.

\subsection{Variation of the normalized irradiance due to the defocus}

\begin{figure}[ht!]
\centering
\includegraphics[trim=0cm 0cm 0cm 0cm, width=0.95\textwidth, clip]{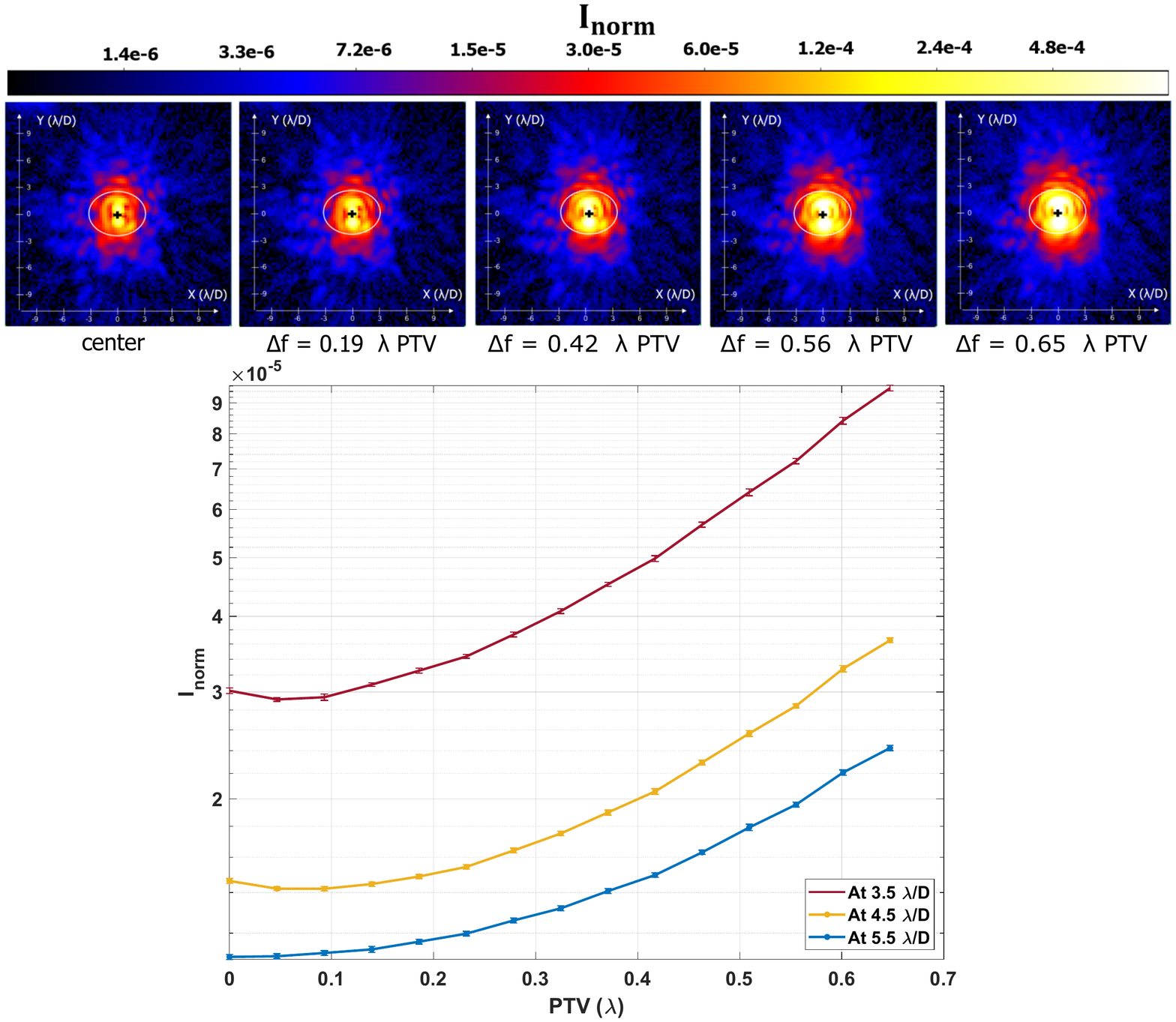}
\caption{The figure shows the normalized irradiance at specific radial locations from the center of the mask: 3.5 $\lambda/D$ (red curve), 4.5 $\lambda/D$ (yellow curve), and 5.5 $\lambda/D$ (blue curve) expressed in terms of $\lambda$ PTV.} \label{fig:EvWaCo_fig17}
\end{figure}

The variation of the normalized irradiance due to defocus is done by translating the star source along the optical axis and measuring the normalized irradiance value. This displacement is expressed in terms of the PTV (peak-to-valley) wavefront error. The PTV wavefront error induced by each displacement is determined by making paraxial calculations using Zemax. Figure \ref{fig:EvWaCo_fig17} (left panel) shows the wavefront error in terms of $\lambda$ PTV induced for each displacement. For each displacement, 101 images are acquired and the normalized irradiance values are calculated over a circular annulus with an extension of 1 $ \lambda/D$. The standard deviation of these values calculated for the 101 images is represented as error bars. At 3 $\lambda$/D from the center of the mask, the normalized irradiance values are better than $10^{-4}$ for a wavefront less than 0.65 $\lambda$ PTV as shown in Figure \ref{fig:EvWaCo_fig17} (right panel).

\subsection{Variation of the normalized irradiance due to Lyot stop misalignment}

\begin{figure}[p]
\centering
\includegraphics[trim=0.5cm 1cm 0.5cm 0.1cm, width=0.95\textwidth, clip]{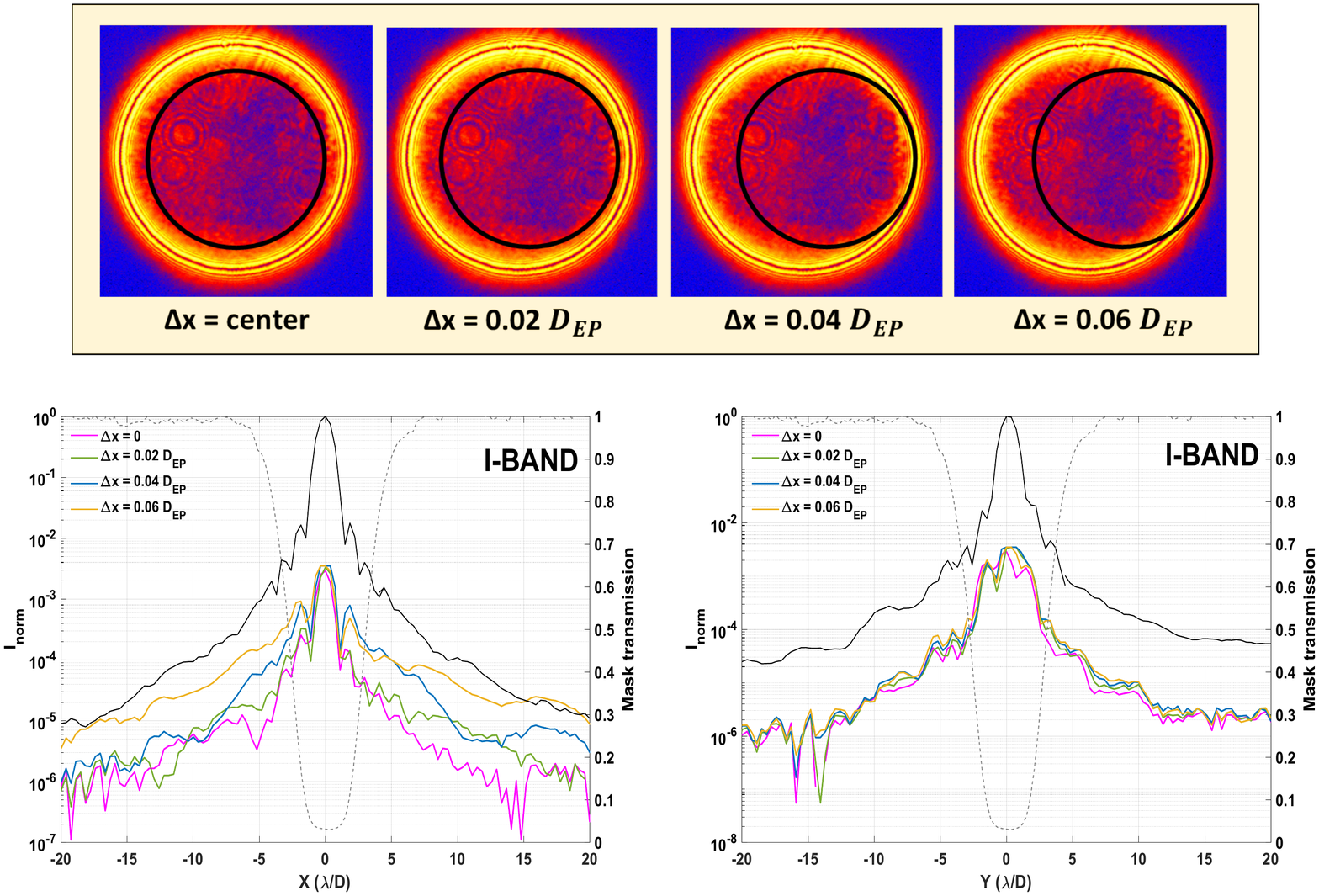}
\caption{The figure at the top shows the position of the Lyot stop with respect to the exit pupil. The figure above shows the normalized irradiance for each corresponding misalignment along the x-axis. We observe that light begins to leak as the Lyot stop is misaligned. However, a misalignment of 74 µm (0.02 $D_{EP}$) does not significantly affect the normalized irradiance at the $IWA_X$ $\approx 4.00$ $\lambda/D$.} \label{fig:EvWaCo_fig18}
\end{figure}

\begin{figure}[p]
\centering
\includegraphics[trim=0.5cm 1cm 0.5cm 0.1cm, width=0.95\textwidth, clip]{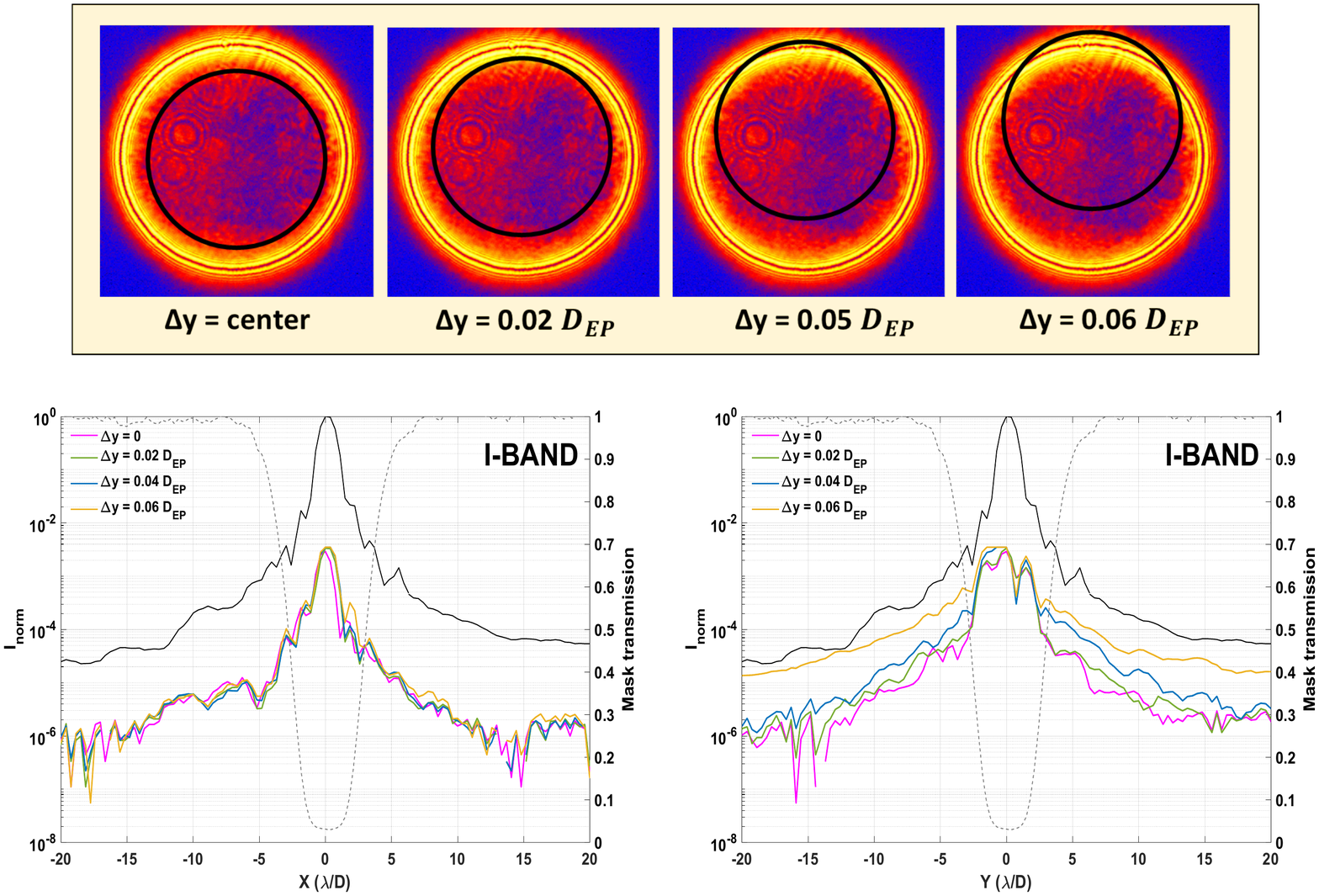}
\caption{The figure at the top shows the position of the Lyot stop with respect to the exit pupil. The figure below shows the normalized irradiance for each corresponding misalignment along the y-axis. We observe that light begins to leak as the Lyot stop is misaligned. However, a misalignment of 84 µm (0.02 $D_{EP}$) does not significantly affect the normalized irradiance at the $IWA_Y$ $\approx$ 3.00 $\lambda/D$.} \label{fig:EvWaCo_fig19}
\end{figure}

The purpose of the Lyot stop is to reduce, if not eliminate, the residual starlight in the exit pupil plane that is due to diffraction. Thus, if this Lyot stop is not properly centered, it causes the star light to leak, thus, degrading the performance of the coronagraph. In this section, we intentionally misaligned the Lyot stop along the x- and the y-directions, and measured the normalized irradiance. Figures \ref{fig:EvWaCo_fig18} and \ref{fig:EvWaCo_fig19} show how the contrast varies when the Lyot stop is not exactly positioned at the center of the exit pupil (EP). The position of the Lyot stop with respect to the center of the exit pupil is marked by the black circle. 

Figure \ref{fig:EvWaCo_fig18} reveals that a misalignment of 74 µm (0.02 $D_{EP}$) along the x-direction does not significantly impact the normalized irradiance at the location of the IWA ($IWA_X$ $\approx$ 4.00 $\lambda/D$, $IWA_Y$ $\approx$ 3.00 $\lambda/D)$. The same phenomenon is observed for a misalignment of 84 µm (0.02 $D_{EP}$) along the y-direction as shown in Figure \ref{fig:EvWaCo_fig19}. These results show the possibility of enlarging the Lyot stop to 90\% the size of the pupil diameter (currently, it is at 81\%) at the expense of a tighter tolerance of the Lyot stop alignment. 

\clearpage

\section{Discussion and Conclusion} \label{sec:discussion}

Our recent results demonstrate that EvWaCo has now entered the regime of small-angle coronagraphy (IWA $\leq$ 4 $\lambda$/D) \citep{Mawet} over a wide spectral band [550 nm – 900 nm]. It is important to note that the results have been obtained by using a setup on a passive support without isolation, in a non-vacuum laboratory environment, and without adaptive optics to correct the wavefront distortions induced by surface errors of the optical components and local air turbulence.
These results show a clear improvement to the performance we have previously reported \citep{Buisset1} in terms of the raw contrast, the spectral band, and the IWA. First, the spectral band has been enlarged from a range of [700 nm, 900 nm] to [550 nm, 900 nm]. Second, the IWA is now fully tunable between 3 $\lambda$/D and 7 $\lambda$/D that is unique. We emphasize that the current IWA has been reduced by a factor of 2 compared to the previous one which was at 6 $\lambda$/D. Finally, we have shown that this mask is able to reach a raw contrast close to a few $10^{-4}$ at 3 $\lambda$/D over the full I-band and at 4 $\lambda$/D over the full R-band. A series of contrast measurements over 8 months confirms its stability and the repeatability of the measurements. In terms of misalignment, the normalized irradiance measured at a distance close to 4 Airy radii is better than $10^{-4}$ even for tilt angles as large as 1 $\lambda/D$ and defocus (0.65 $\lambda$ PTV) at the IWA. The Lyot stop can be moved by at most $\pm$ 74 µm without affecting the normalized irradiance that is also compatible with common mechanical tolerances. Finally, we have shown that a companion, 25000 times fainter than the star, can be detected with a good SNR at 5.5 $\lambda/D$ from the star center.

The results presented in this paper show that the best contrast performance is reached for the minimum contact area that provides a fully achromatic quasi-gaussian mask transmission. Therefore, the objective of the FPM future upgrades will consist of improving the mask opto-mechanical design to i) reduce the size of this contact to a negligible value, ii) keep the current performance and stability, and iii) increase the maximum available mask size.

The future development of EvWaCo will aim at developing a full prototype that will include an adaptive optics and demonstrating on-sky performance. The full prototype will demonstrate the capability to reach contrasts that are in line with the laboratory results during ground-based observations. The first step will consist of installing an EvWaCo prototype on the 2.4 m Thai National Telescope at horizon 2022.

\newpage
\section{Acknowledgment}
This work is carried out at the NARIT Center for Optics and Photonics. We thank the mechanical engineers of the Operations Department of NARIT for the fabrication of optical component supports in the testbed.

The Thai Space Consortium (TSC) is an association composed of the National Astronomical Research Institute of Thailand (NARIT), Siam Photon, Geo-Informatics and Space Technology Development Agency (GISTDA), Suranaree University of Technology (SUT), King Mongkut's University of Technology North Bangkok (KMUTNB), King Mongkut's Institute of Technology Ladkrabang (KMITL), and National Innovation Agency (NIA). TSC is funded by PMU-B, Office of National Higher Education Science Research and Innovation Policy Council. The research is supported by the Program Management Unit for Human Resources \& Institutional Development, Research and Innovation, NXPO (Grant number: B05F630115). We also thank the Chulalongkorn University's CUniverse (CUUAASC) grant for the support.

Finally, we are very grateful for the invaluable insights from the reviewer that helped us improve the quality of our paper.
\newpage
\section{Appendix}
\subsection{Data Acquisition Procedure}

In the experiment, we obtained two sets of off-axis PSF images, $PSF_{coro} (x,x_o,\lambda)$, placed at a distance of $x_o \approx 30 \lambda/D$. On the first set, we installed an optical density OD2 in front of the detector with the following transmission on each bandpass: $t_{I-band}$ = $2.10^{-3}$ and $t_{R-band}$ = $1.3.10^{-3}$. We adjusted the integration time so that the peak of the star PSF is close to the maximum dynamic range of the camera and recorded N = 101 frames of the same integration time. Then, we recorded 11 dark images and 11 bias images at an integration time equal to 1 s and 0 s, respectively. We calculated the master dark frame and the master bias frame as the median of the 11 single dark frames and 11 bias frames, respectively. We subtracted the master bias and the dark frames to each single frame of the off-axis PSF and calculated the median of the 101 images. We measured on this median image the peak value $S_{Star,peak}$ of the (unsaturated) off axis PSF. 

On the second set, we removed OD2 and measured the signal $S_{Star}$ (x,y) of the PSF with the same integration time $\tau$, same number N of images and identical pre-processing method. This image of the off-axis PSF has now a saturated core but a high Signal to Noise Ratio (SNR) in the PSF wings.

To obtain  $PSF_{coro} (x, 0 ,\lambda)$, we centered the PSF on the occulting mask and recorded the signal $S_{Residual}$ (x,y) at the same integration time $\tau$. Thus, the normalized irradiance is calculated using Equation \ref{eqn2}.

\begin{equation}
I_{norm}(x_o, \lambda) = \frac{S_{Residual}T_{OD2}}{S_{Star,Peak}}
\label{eqn2}	
\end{equation}

To calculate the raw contrast, we divided $I_{norm}$ by the mask transmission obtained from the experiment. For example, from Figure 10, at 3 $\lambda/D$, the normalized irradiance is approximately at $I_{norm} \approx 1.10^{-4}$ in the I-band. Thus, the corresponding raw contrast at this location is equal to $2.10^{-4}$.

\end{document}